\documentclass[aps,pre,twocolumn,superscriptaddress]{revtex4}
\pdfoutput=1
\usepackage{graphicx}
\usepackage{amsmath}
\usepackage{amssymb}
\usepackage{color}
\usepackage{bm}

\definecolor{red}{rgb}{0.75,0,0}
\definecolor{blue}{rgb}{0,0,0.75}
\definecolor{green}{rgb}{0,0.5,0}

\usepackage[pdftex, pdfborder={0 0 0}, colorlinks=true, linkcolor=red, citecolor=blue, urlcolor=green]{hyperref}

\newcommand{\taua}{\tau_\text{a}}
\newcommand{\taup}{\tau_\text{p}}
\newcommand{\sigmap}{\bm{\sigma_\mathbf{p}}}

\DeclareMathOperator{\tr}{tr}

\begin{document}	
	
\title{Banding, Excitability and Chaos in Active Nematic Suspensions}

\author{L. Giomi}
\email{lgiomi@seas.harvard.edu}
\affiliation{School of Engineering and Applied Sciences, Harvard University, Cambridge, MA 02138,  USA}

\author{L. Mahadevan}
\affiliation{School of Engineering and Applied Sciences, Harvard University, Cambridge, MA 02138,  USA}
\affiliation{Department of Physics, Harvard University, Cambridge, MA 02138,  USA}

\author{B. Chakraborty}
\affiliation{Martin A. Fisher School of Physics, Brandeis University, Waltham, MA 02454, USA}

\author{M. F. Hagan}
\email{hagan@brandeis.edu}
\affiliation{Martin A. Fisher School of Physics, Brandeis University, Waltham, MA 02454, USA}
\date{\today}

\begin{abstract}
\noindent Motivated by the observation of highly unstable flowing states in suspensions of microtubules and kinesin, we analyze a model of mutually-propelled filaments suspended in a solvent. The system undergoes a mean-field isotropic-nematic transition for large enough filament concentrations when the nematic order parameter is allowed to vary in space and time. We analyze the model in two contexts: a quasi-one-dimensional channel with no-slip walls and a two-dimensional box with periodic boundaries. Using stability analysis and numerical calculations we show that the interplay between non-uniform nematic order, activity, and flow results in a variety of complex scenarios that include spontaneous banded laminar flow, relaxation oscillations, and chaos.
\end{abstract}

\maketitle
\section{\label{sec:introduction}Introduction}

\noindent Active hydrodynamics describes the collective motion of microscopic particles constantly maintained out of equilibrium by internal energy sources. Colonies of swarming bacteria, \emph{in vitro} mixtures of cytoskeletal filaments and motor proteins and vibrated granular rods are common examples of active systems and now \emph{active} has become standard terminology for any system whose constituents drive themselves mechanically by extracting and dissipating energy from their environment. Originating from pioneering works by Pedley and Kessler \cite{Pedley:1992},  Vicsek \emph{et al}. \cite{Vicsek:1995} and Simha and Ramaswamy \cite{Simha:2002}, active matter research has  blossomed  to encompass diverse systems  and scales ranging from  animal groups to subcellular matter \cite{Ramaswamy:2010}.

Because active particles typically have elongated shapes, their collective behavior has been often described using the language of liquid crystals \cite{Gruler:1999,Kruse:2004}. In this regard, an important distinction among active particles concerns the possibility of forming phases characterized by nematic or polar order. While all elongated particles can form a nematic phase at sufficient densities and levels of activity, particles which have an asymmetry associated with their mutual interaction can additionally form a phase characterized by a non-zero macroscopic polarization. Active particles can also be distinguished in terms of their locomotion characteristics: self-propelled particles (SPP) are endowed with an internal engine and, typically but not necessarily, with appendages that allow them to swim in a fluid or crawl on a substrate. For example, bacteria \cite{Darnton:2010}, large animals such as fish or birds \cite{Ballerini:2008}, and catalytic motors \cite{Paxton:2004} belong in this category. Cytoskeletal filaments, on the other hand, cannot propel themselves, but move in a solvent through the action of motor proteins, which are themselves powered by the hydrolysis of adenosine triphosphate (ATP). Bundles of molecular motors attach to pairs of filaments and, during an ATP cycle, slide the filaments with respect to each other. We will refer to this type of active elements as mutually-propelled particles (MPP). There is finally a third class of systems in which activity is provided through vibration. Vertically shaken granular rods, for instance, gain and dissipate energy while bouncing on a substrate, resulting in a two-dimensional motion along their major axis \cite{Narayan:2007,Deseigne:2010}.

Most theoretical effort in modeling active systems characterized by liquid crystalline order has focused on constructing hydrodynamic equations that, in addition to the usual liquid crystalline elasticity, can account for the additional forces and currents originated from the activity. This task has been achieved by incorporating phenomenological non-equilibrium terms in the hydrodynamic equations of nematic and polar liquid crystals \cite{Simha:2002,Kruse:2004,Voituriez:2005} or by applying the tools of non-equilibrium statistical mechanics to specific microscopic models \cite{Liverpool:2003,Ahmadi:2006,Baskaran:2008,Baskaran:2009}. This program has generated a variety of predictions, which include the existence of \emph{giant} density fluctuations in active nematics \cite{Ramaswamy:2003,Mishra:2006,Narayan:2007}, spontaneously flowing states \cite{Voituriez:2005,Marenduzzo:2007,Giomi:2008,Edwards:2009,Chate:2006,Ginelli:2010,Fielding:2011,Marchetti:2011}, unconventional rheological properties \cite{Fielding:2011,Cates:2008,Sokolov:2009,Lau:2009,Giomi:2010,Saintillan:2010} and a plethora of a novel hydrodynamic instabilities with no counterpart in passive complex fluids \cite{Simha:2002,Ramaswamy:2007,Saintillan:2007,Saintillan:2008,Sankararaman:2009,Baskaran:2009}; a recent overview can be found in \cite{Ramaswamy:2010}.

In spite of the vast theoretical work on \emph{active liquid crystals}, little consideration has been given to the possibility of spatial and temporal variations in the order parameter; a recent exception is the work of Mishra \emph{et al}. \cite{Mishra:2010} who considered self-propelled polar rods moving on a frictional substrate, with a density-driven mean field transition from the isotropic to the polar phase and showed that when the self-propulsion velocity exceeds a threshold value, the uniformly polarized moving state becomes unstable to spatial fluctuations which organize into stripes of different density and polarization.

Here we consider the case of an active nematic suspension motivated by the observation of highly unstable flowing states in assemblies of microtubules and kinesin \cite{Sanchez:2010}, a model for mutually-propelled elongated particles in a solvent. The system undergoes a mean-field isotropic-nematic transition for large enough filament concentrations and the nematic order parameter is allowed to vary in space and time. We use  stability analysis and numerical simulations to analyze the model in two geometries: a quasi-one-dimensional channel with no-slip walls and a two-dimensional box with periodic boundaries. In the channel geometry, moderate activity levels lead to spontaneous laminar flow, as seen in earlier works \cite{Voituriez:2005,Giomi:2008} that assumed a constant magnitude of the nematic order parameter. Upon increasing the activity past a threshold value, however, fluctuations in magnitude of the nematic order parameter leads to oscillatory flow in which the nematic director periodically switches orientation. In the two-dimensional box, the interplay between non-uniform nematic order, activity, and flow results in a variety of complex scenarios that include spontaneous laminar flow,  relaxation oscillations reminiscent of excitable media, and chaos. A detailed analysis allows us to uncover the origin of oscillations in the system and characterize the chaotic regime, wherein we see behavior consistent with turbulent flow even in the low Reynolds number regime, expanding on and complementing a recent short report of some of our findings \cite{Giomi:2011}.

This article is organized as it follows. In Sec. \ref{sec:hydrodynamics} we introduce the hydrodynamic equations for an active suspension of mutually propelled filaments. In Sec. \ref{sec:channel} we analyze the equations for a quasi-one-dimensional channel of infinite length and finite width endowed with no-slip walls. In Sec. \ref{sec:plane} we consider an active nematic suspension in a two-dimensional container with periodic boundaries. We then present a minimal model that demonstrates oscillatory behavior, and characterize the the chaotic regime. Finally, we present our conclusions in Sec. \ref{sec:conclusions}.

\section{\label{sec:hydrodynamics}Hydrodynamical equations of motion}

A fluid of orientable fore-aft symmetric particles can generally exist in two phases: isotropic (I) and nematic (N). In the latter phase, the particles are orientationally ordered with an average orientation characterized by the nematic director field ${\bf n}$. For microscopic particles in suspension, such as colloidal rods or biological filaments, the IN transition is driven by density: when the concentration of particles overcomes some critical value $c^{*}$, the particles form a nematic phase in order to maximize entropy. In a two-dimensional \emph{equilibrium} fluid of slender rods the critical concentration is given by $c^{*}=3\pi/2\ell^{2}$ where $\ell$ is the length of the rods \cite{Kayser:1978} and the phase transition is of the Kosterlitz-Thouless type. The anisotropy of a nematic phase is expressed through the nematic tensor $Q_{ij}$ \cite{DeGennesProst}, which for uniaxial nematics reads:
\begin{equation}
Q_{ij} = S\left(n_{i}n_{j}-\frac{1}{d}\,\delta_{ij}\right)	
\end{equation}
The nematic tensor $Q_{ij}$ is by construction traceless and symmetric, thus in $d=2$ it consists of only two independent degrees of freedom. The nematic phase has orientational order ($S\ne 0$) and is invariant under inversion of the director field: ${\bf n}\rightarrow-{\bf n}$. Here the extent of nematic alignment is expressed in terms of a scalar nematic order-parameter $S$:
\begin{equation}
S = \frac{1}{d-1}\,\langle\,d\,|{\bf a}\cdot{\bf n}|^{2}-1\rangle\,,
\end{equation}
where ${\bf a}$ is the axis of the molecules, $d$ is the dimension of the system and the angular brackets denote a thermal average. In a suspension of rod-like particles, $S$ depends on the local concentration of the particles and, in equilibrium passive systems, is constant across the sample since diffusion drives the fluid toward a homogeneous state. In active systems, however, activity can build up density inhomogeneities and the order parameter may exhibit spatial fluctuations. Moreover, since the effects of activity are generally enhanced by local orientational order, coupling between order, activity and flow can amplify these fluctuations. In the following we describe a set of hydrodynamic equations suitable to describe a suspension of active particles whose nematic order is allowed to vary in space and time as a consequence of activity and flow.

Let us consider a concentration $c$ of rod-like active particles of length $\ell$ and mass $M$ suspended in a solvent of concentration $\rho_{\rm solvent}$. The total density of the system $\rho=Mc+\rho_{\rm solvent}$ is conserved and the fluid is incompressible. Since the total number of particles is also constant, the concentration $c$ obeys a continuity equation of the form:
\begin{equation}\label{eq:continuity-equation}
\partial_{t}c = -\nabla\cdot[c({\bf v}+{\bf v}^{a})-\bm{D}\nabla c]\,,
\end{equation}
where ${\bf v}$ is the bulk flow velocity, ${\bf v}^{a}$ is the velocity at which the particles actively move relative to the flow, and $\bm{D}$ is the diffusion tensor, which  two-dimensional uniaxial nematics reads:
\begin{equation}
D_{ij} = D_{0}\delta_{ij}+D_{1}Q_{ij}\,,
\end{equation}
where $D_{0}=(D_{\parallel}+D_{\perp})/2$, $D_{1}=D_{\parallel}-D_{\perp}$ and $D_{\parallel}$ and $D_{\perp}$ are respectively the bare diffusion coefficients along the parallel and perpendicular directions of the director field. The active current ${\bf j}^{a}=c{\bf v}^{a}$ has been modeled in different ways and depends on whether the system is in a nematic or polar phase. In polar systems, active particles are collectively propelled in the direction of the macroscopic polarization ${\bf P}$; thus ${\bf v}^{a}=v_{0}{\bf P}$ with $v_{0}$ the average velocity of an individual active particle. Thus, for example,  for bacterial suspensions $v_{0}$ is constant and represents the average swimming velocity of an individual bacterium \cite{Pedley:1992}, while for mutually propelled particles, such as cytoskeletal filaments pushing against each other through the action of motor clusters, $v_{0}$ depends on the average concentration of the filaments. Thus $v_{0}=\alpha_{1}c$ with $\alpha_{1}=u_{0}\ell^{2}$, where $u_{0}$ is the propulsion velocity for unit concentration and is proportional to the rate of ATP consumption. This leads to an active current of the form ${\bf j}^{a}=\alpha_{1}c^{2}{\bf P}$ \cite{Ahmadi:2006,Liverpool:2007}.

\begin{figure}
\centering
\includegraphics[width=1\columnwidth]{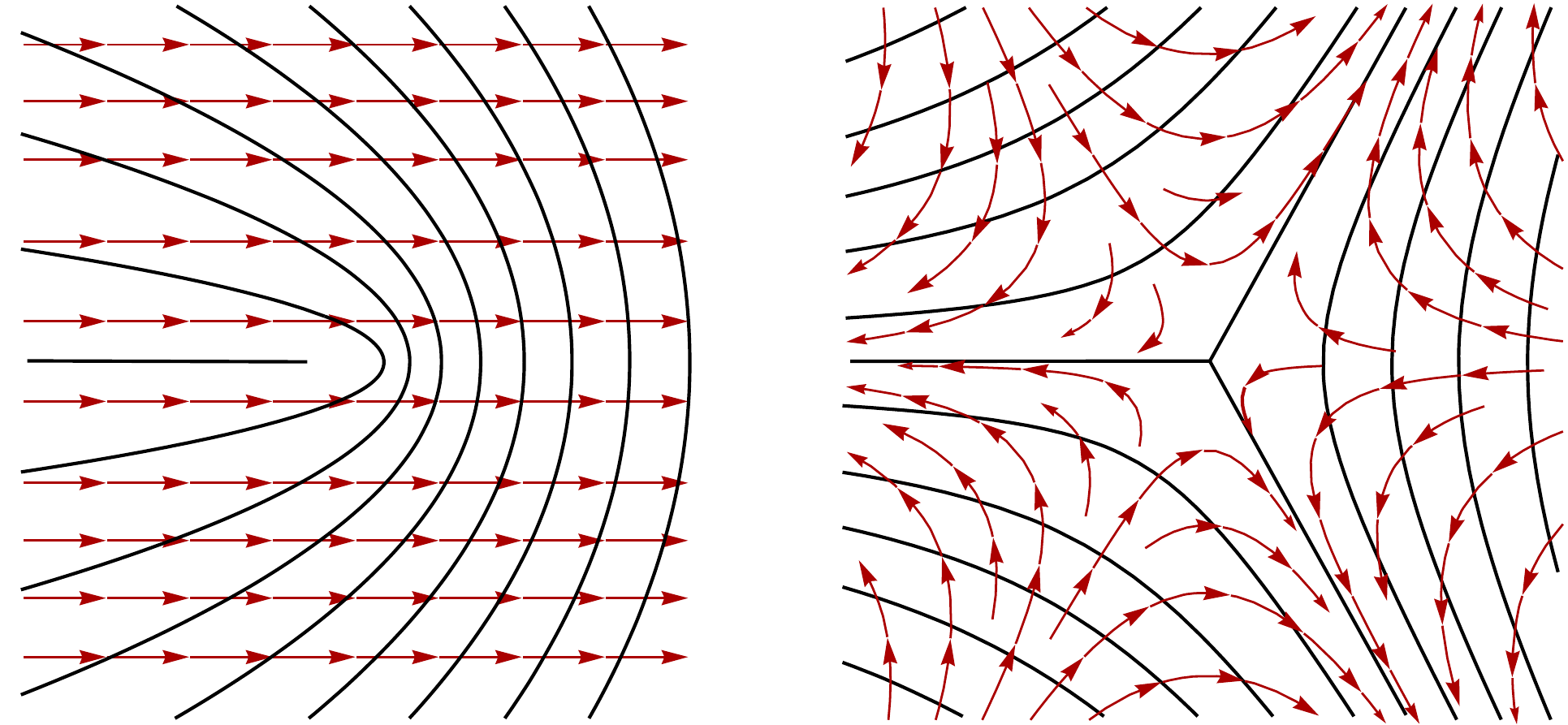}	
\caption{\label{fig:active-current}An example of the active currents resulting from Eq. \eqref{eq:active-current} in the presence of large distortions of the director field, such as those which occur near a disclination. As noted in Ref. \cite{Narayan:2007}, the tilt in the director field surrounding a $+1/2$ disclination (left) results in a collective drift of particles in the direction indicated by the red arrows (i.e. toward the ``nose'' of the defect). On the other hand, a $-1/2$ disclination (right) will produce the same amount of incoming and outgoing currents and thus zero net flux.}
\end{figure}

For a nematic suspension, on the other hand, the active particles move along ${\bf n}$ and $-{\bf n}$ at the same rate; thus if the director field is uniform across the system, there will be no net flux of particles across an arbitrarily small domain and ${\bf j}^{a}={\bf 0}$. However, in the presence of a non-uniform director field or equivalently a non-uniform nematic order parameter, there will be regions of fluid moving faster than others and thus a current. Such a current must depend on the derivatives of the nematic tensor rather than on $Q_{ij}$ itself. The simplest term of this type with the correct tensorial structure is given by $v_{i}^{a}=v_{0}\ell\partial_{j}Q_{ij}$, which for mutually propelled particles gives a current:\
\begin{equation}\label{eq:active-current}
j_{i}^{a}=-\alpha_{1}c^{2}\partial_{j}Q_{ij}\,,	
\end{equation}
where $\alpha_{1}$ is a constant with dimensions of inverse time. The negative sign in Eq. \eqref{eq:active-current} reflects the fact that the flux of active particles is directed from regions populated by fast moving particles to regions of slow moving particles. The active current ${\bf j}^{a}$ has been derived in the form \eqref{eq:active-current} by Ahmadi \emph{et al}. starting from a microscopic model of filaments interacting through a motor cluster \cite{Ahmadi:2006} and later by Lau and Lubensky for swimming bacteria \cite{Lau:2009} (the dependence on the concentration $c$ is different in the latter case because of the different microscopic model). Ramaswamy and coworkers argued that such active currents are responsible for the existence of \emph{giant} fluctuations in the number density of active nematic particles in the presence of noise \cite{Ramaswamy:2003,Narayan:2007}. Since spatial variations of the order parameter were neglected in that work, the only driving force for active currents arose from curvature (i.e. tilt) in the director field orientation, hence the name ``curvature-induced currents'' coined in \cite{Ramaswamy:2003}. Here we show that a more complete description, in which both the orientation of the director field and the nematic order parameter are allowed to vary, leads to additional complex phenomena.

Next we construct a set of hydrodynamic equations for the nematic tensor $Q_{ij}$. These can be written in the generic form:
\begin{equation}\label{eq:q-generic}
[\partial_{t}+{\bf v}\cdot\nabla]Q_{ij} = \Omega_{ij}^{(r)}+\Omega_{ij}^{(v)}+\Omega_{ij}^{(a)}\,,
\end{equation}
where the rates $\Omega_{ij}^{(r)}$, $\Omega_{ij}^{(v)}$ and $\Omega_{ij}^{(a)}$ embody respectively the relaxational dynamics, the coupling with the flow, and the active contribution to the dynamics of the nematic tensor. Following Olmsted and Goldbart \cite{Olmsted:1992}, the rates on the right-hand side of Eq. \eqref{eq:q-generic} can be obtained phenomenologically by constructing all possible traceless-symmetric combinations of the relevant fields of the theory. These are the strain-rate tensor $u_{ij}=\frac{1}{2}(\partial_{i}v_{j}+\partial_{j}v_{i})$, the vorticity tensor $\omega_{ij}=\frac{1}{2}(\partial_{i}v_{j}-\partial_{j}v_{i})$, and the molecular tensor $H_{ij}=-\delta F/\delta Q_{ij}$ defined from the two-dimensional Landau-de Gennes free energy $F$ \cite{DeGennesProst}. In two dimensions this reads \footnote{In three dimensions, on the other hand, the Landau-de Gennes free energy contains an extra cubic term of the form $\frac{1}{3}B\,Q_{ij}Q_{jk}Q_{ki}$ that allows the mean-field transition to be first-order.}:
\begin{equation}
F = \int dA\,[\tfrac{1}{2}A\,Q_{ij}Q_{ij}+\tfrac{1}{4}C\,(Q_{ij}Q_{ij})^{2}+\tfrac{1}{2}K\,\partial_{i}Q_{jk}\partial_{i}Q_{jk}]
\end{equation}
where the coefficients $A$ and $C$  characterize the location of a second order phase transition. Since $\tr(\bm{Q}^{2})=S^{2}/2$, at equilibrium one has that $S=\sqrt{-2A/C}$. In a hard-rod fluids, when the IN transition is driven only by the concentration of the nematogens, one can chose for instance: $A/K=(c^{*}-c)/2$ and $C/K=c$, so that:
\begin{equation}
S = \sqrt{1-c^{*}/c}.
\end{equation}
Thus for $c\gg c^{*}$, $S\sim 1$ while for $c<c^{*}$, $S=0$. The last term in the free energy expression is the Frank elastic energy in the one elastic constant approximation \cite{DeGennesProst}. As for passive liquid crystals, the relaxational dynamics of $Q_{ij}$ is driven by the molecular tensor $H_{ij}$:
\begin{equation}\label{eq:omega-r}
\Omega_{ij}^{(r)} = \gamma^{-1}H_{ij} = -\gamma^{-1}[(A+\tfrac{1}{2}S^{2}C)Q_{ij}-K\Delta Q_{ij}]
\end{equation}
where $\gamma$ is a type of rotational viscosity. The coupling between nematic order and flow is found by constructing all possible symmetric traceless tensors from the products of $u_{ij}$, $\omega_{ij}$, and $Q_{ij}$ \cite{Olmsted:1992}. This yields:
\begin{gather}
\Omega_{ij}^{(v)}
= \beta_{1}u_{ij}+\beta_{2}(\omega_{ik}Q_{kj}-Q_{ik}\omega_{kj})\notag\\
+\beta_{3}\left[u_{ik}Q_{kj}+Q_{ik}u_{kj}-\frac{2}{d}\,\tr(\bm{u}\bm{Q})\,\delta_{ij}\right]
\end{gather}
In two dimensions the last term is identically zero and $\Omega_{ij}^{(v)}$ simplifies to:
\begin{equation}\label{eq:omega-v}
\Omega_{ij}^{(v)} = \beta_{1}u_{ij}+\beta_{2}(\omega_{ik}Q_{kj}-Q_{ik}\omega_{kj})
\end{equation}
where the coefficients $\beta_{1}$ and $\beta_{2}$ can be found by comparing this expression with the standard Ericksen-Leslie theory \cite{DeGennesProst,LandauLifshitz}, which gives $\beta_{1}=\lambda S$ and $\beta_{2}=-1$ \cite{Olmsted:1992}. Here $\lambda$ is the flow-aligning parameter which dictates how the director field rotates in a shear flow. In passive liquid crystals, for $|\lambda|>1$, the director tends to align to the flow direction at an angle $\theta_{0}$ such that $\cos 2\theta_{0}=1/\lambda$, while for $|\lambda|<1$, it forms rolls across the system. These regimes are known as ``flow aligning'' and ``flow tumbling'' respectively. The value of $\lambda$  has even greater significance in active systems; together with the magnitude of forces exerted by the active particles it dramatically influences the flow behavior and rheological properties of the system \cite{Giomi:2008,Giomi:2010}.

The active contribution to the dynamics of the nematic tensor was derived by Ahamadi \emph{et al}. \cite{Ahmadi:2006} and is directly proportional to the nematic tensor
\begin{equation}\label{eq:omega-a}
\Omega_{ij}^{(a)}=\alpha_{0}Q_{ij}\,.
\end{equation}
However, given the structure of Eqs. \eqref{eq:q-generic} and \eqref{eq:omega-r}, such a term can be simply incorporated into the molecular field, leading to a redefinition of the critical concentration $c^{*}$. Here we will drop this explicit dependence for sake of brevity, but remember that that the critical concentration $c^{*}$ associated with the IN transition does depend on the activity. The corresponding reactive stress tensor can be obtained from Eqs. \eqref{eq:q-generic}, \eqref{eq:omega-r} and \eqref{eq:omega-v} using the standard energy conservation and entropy production argument (see, for example, Ref. \cite{LandauLifshitz}). This gives, after some algebra:
\begin{equation}\label{eq:sigma-e}
\sigma_{ij}^{(e)} = -\lambda S H_{ij}+Q_{ik}H_{kj}-H_{ik}Q_{kj}
\end{equation}
Finally the flow velocity obeys the Navier-Stokes equation, with the total stress tensor given by:
\begin{equation}\label{eq:sigma-tot}
\sigma_{ij} = 2\eta u_{ij}-p\delta_{ij}+\sigma_{ij}^{(e)}+\alpha_{2}c^{2}Q_{ij}
\end{equation}
where $\eta$ is fluid viscosity of the fluid and $p$ the pressure. The last term was established in the seminal work of Pedley and Kessler \cite{Pedley:1992} and represents the tensile/contractile stress exerted by the active particles in the direction of the director field ${\bf n}$. The $c^{2}$ dependence again arises because the propulsion, in our analysis, is provided by pair interactions of the filaments through motors. A detailed derivation can be found in Ref. \cite{Liverpool:2007}.

Summarizing, the hydrodynamics of an incompressible nematic suspension of mutually propelled filaments is governed by the following set of differential equations for the particle concentration $c$, the nematic tensor $Q_{ij}$, and the flow velocity ${\bf v}$ (with components $v_i$):
\begin{gather} \label{eq:fullEquations}
\rho\partial_{t}v_{i} = \eta \partial^2_i v_{i} -\partial_{i}p+\partial_{j}\tau_{ij}  \\[5pt]
[\partial_{t}+{v_i} \partial_i ]c = \partial_{i}[(D_{0}\delta_{ij}+D_{1}Q_{ij})\partial_{j}c+\alpha_{1}c^{2}\partial_{j}Q_{ij}] \nonumber \\[5pt]
[\partial_{t}+v_i \partial_i]Q_{ij} = \lambda S u_{ij}+Q_{ik}\omega_{kj}-\omega_{ik}Q_{kj}+\gamma^{-1}H_{ij} \nonumber
\end{gather}
where we defined $\tau_{ij}=\sigma_{ij}^{(e)}+\alpha_{2}c^{2}Q_{ij}$. Here we have neglected the inertial convective term in the velocity equation because  we are interested in fluids of cytoskeletal filaments and motor proteins for which the typical Reynolds number is small. However, since both the nematic order and the concentration of particles can be advected by the flow, the associated terms in the equations for $c$ and $Q_{ij}$ cannot be neglected.

The dynamics of such an active nematic suspension is governed by the interplay between the active forcing, whose rate $\taua^{-1}$ is proportional to the activity parameters $\alpha_{1}$ and $\alpha_{2}$, and the relaxation of the passive structures, the solvent and the nematic phase, in which energy is dissipated or stored. The response of the passive structures, as described here, occurs at three different time scales: the relaxational time scale of the nematic degrees of freedom $\ell^{2}/(\gamma^{-1}K)$, the diffusive time scale $\ell^{2}/D_{0}$, and the dissipation
time scale of the solvent $\rho L^{2}/\eta$, where $L$ is the system size. While the presence of three dimensionless parameters makes for a very rich phenomenology, for simplicity we choose parameter values in this work so that the three passive time scales are of the same magnitude $\taup$. When $\taua\gg\taup$, the active forcing is irrelevant and the system behaves like a traditional passive suspension. On the other hand, when $\taua\sim\taup$, the passive structures can balance the active forcing leading to a stationary regime in which active stresses are accommodated via both elastic distortion and flow. Finally, when $\taua\ll\taup$ the passive structures respond too slowly to compensate active forces, leading to a dynamical and possibly chaotic interplay between activity, nematic order and flow. In the rest of the paper, we quantify these different regimes.

For further manipulations it is convenient to make the system dimensionless by scaling all lengths using the rod length $\ell$, scaling time with the relaxation time of the director field $\taup=\ell^{2}/(\gamma^{-1}K)$, and scaling stresses  by the elastic stress $\sigma=K\ell^{-2}$.

\section{\label{sec:channel}Channel geometry}

\subsection{Overview}

The simplest geometry in which to analyze the hydrodynamic equations given in the previous section is a two-dimensional channel of infinite length and finite width. This geometry has been studied in detail for active nematic and polar suspensions under the assumption of constant magnitude of the nematic order parameter \cite{Voituriez:2005,Marenduzzo:2007,Giomi:2008,Edwards:2009,Giomi:2010}. The most striking feature of active nematic fluids in a channel is the ``spontaneous flow transition'': when the activity parameter $\alpha_{2}$ is increased past a threshold, the system goes from a stationary state in which the director field is parallel to the walls of the channel  to a state of non-uniform orientation and flow. Here we show that lifting the assumption of constant nematic order parameter leads to a second transition to oscillatory flow not considered previously in theories of active nematics.

We consider a channel of infinite length along the $x$ direction of a Cartesian frame and finite width $L$ along $y$. The channel is bounded by no-slip surfaces at $y=0$ and $y=L$. Assuming translational invariance in the $x$ direction, the flow field is completely defined by the velocity field $v_x= v_x(y), v_y=0$ since the incompressibility condition implies that
\begin{equation}
\nabla \cdot {\bf v} = \partial_{y} v_{y} = 0;
\end{equation}
thus, $v_{y}=0$ since the fluid is confined. The strain rate tensor has only one non-zero component $u_{xy}=\partial_{y}v_{x}/2 \equiv u/2$. Calling $\theta$ the angle between the director field ${\bf n}$ and the $x$ axis, the nematic tensor can be expressed in the simple form:
\begin{equation}
\bm{Q} =
\frac{S}{2}
\left(
\begin{array}{cc}
\cos 2\theta &\phantom{-}\sin 2\theta \\
\sin 2\theta & -\cos 2\theta	
\end{array}
\right)
\end{equation}
Under these conditions the hydrodynamic equations (\ref{eq:fullEquations}) simplify to:
\begin{gather}
\rho\,\partial_{t}v_{x}  =  \partial_{y}\sigma_{xy} \nonumber \\[5pt]
\partial_{t}c  = \partial_{y}\left[\left(D_{0}-\tfrac{1}{2}D_{1}S^{2}\cos 2\theta\right)\partial_{y}c-\tfrac{1}{2}\alpha_{1}c^{2}\partial_{y}(S\cos 2\theta)\right] \nonumber \\[5pt]
\partial_{t}S = \partial_{y}^{2}S-S[A(c)+\tfrac{1}{2}C(c)S^{2}-\lambda u \sin 2\theta+4(\partial_{y}\theta)^{2}] \nonumber \\[5pt]
\partial_{t}\theta = \partial_{y}^{2}\theta+2S^{-1}\partial_{y}S\,\partial_{y}\theta-\tfrac{1}{2} u (1-\lambda\cos 2\theta)
\label{eq:slab-geometry}
\end{gather}
where we have assumed that the inertia of the active particles is negligible and the dependence of the coefficients $A$ and $C$ on the concentration $c$ is as explained previously (The procedure for decoupling the angle $\theta$ and the order parameter $S$ is given in Appendix \ref{sec:appendix1}). Furthermore, the total shear stress is finally given by:
\begin{gather}
\sigma_{xy}
= \eta u + \tfrac{1}{2}\alpha_{2}c^{2} S\sin 2\theta
+ \tfrac{1}{2}S\lambda\sin 2\theta\,\partial_{y}^{2}S\notag\\[5pt]
+ 2S(1-\lambda\cos 2\theta)\partial_{y}S\,\partial_{y}\theta
+ 2S(1-\lambda\cos 2\theta)\partial_{y}^{2}\theta\notag\\[5pt]
+ S\lambda\sin 2\theta[A(c)+\tfrac{1}{2}C(c)S^{2}+4(\partial_{y}\theta)^{2}]
\label{eq:sxy}.
\end{gather}
To complete the formulation of the problem, we need to specify some boundary conditions. Here we assume that
\begin{gather}
v_{x}(0)=v_{x}(L)=0,~\theta(0)=\theta(L)=0, \nonumber\\ ~c'(0)=c'(L)=0,~S'(0)=S'(L)=0. \label{bc}
\end{gather}
Here the condition on $\theta$ assumes strong anchoring with the filaments parallel to the wall at the boundaries, and the conditions on $c$ and $S$ imply that there is no current flowing through the walls. In particular, from Eq. \eqref{eq:continuity-equation} and \eqref{eq:active-current}:
\begin{equation}
j_{y} = \left(D_{0}-\tfrac{1}{2}D_{1}S^{2}\cos 2\theta\right)\partial_{y}c-\tfrac{1}{2}\alpha_{1}c^{2}\partial_{y}(S\cos 2\theta)\,.
\end{equation}
The condition $j_{y}(0)=j_{y}(L)=0$ requires $c'=0$ and $S'=0$ at the boundaries.  As initial conditions we take $c=c_{0}$ (constant), $v_{x}=0$ and $\theta$ and $S$ randomly distributed.

\begin{figure}
\centering
\includegraphics[width=1.\columnwidth]{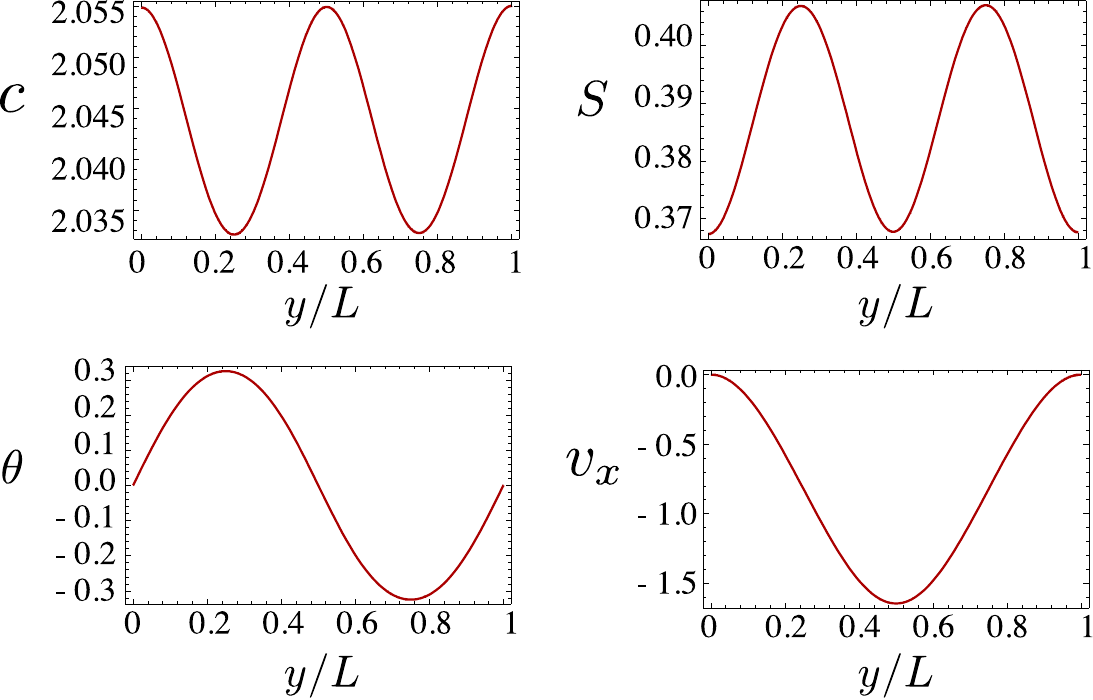}
\caption{\label{fig:spontaneous-flow}The hydrodynamic fields $c$, $S$, $\theta$ and $v_{x}$ as a function of $y/L$ in the spontaneously flowing regime for the channel geometry obtained by solving Eqs. \eqref{eq:slab-geometry} with boundary conditions \eqref{bc}. The initial concentration of the active particles is set to $c_{0}=2c^{*}$, $\alpha_{2}=1.5$ and $\alpha_{1}=0.1\alpha_{2}$. The other parameters are $\lambda=0.1$, $L=5$ and $\eta=D_{0}=D_{1}=1$. For these values, the spontaneous flow transition occurs at $\alpha_{2}=1.24069$.}	
\end{figure}

We solved the hydrodynamical equations (\ref{eq:slab-geometry}-\ref{eq:sxy}) with the boundary conditions (\ref{bc}) numerically after dropping all derivatives of order higher than two in the equation for $v_{x}$ to avoid the use of fictitious boundary conditions. In all our numerical calculations we set $\alpha_{1}=0.1\alpha_{2}$, although other parameters are changed as indicated. Our simulations show that the system exhibits three different regimes determined by the values of the activity parameter $\alpha_{2}$ and the flow-alignment parameter $\lambda$. For small activity, the homogeneous stationary state is the only stable solution, with:
\begin{equation}
v_{x}=0\,,\quad
c = c_{0}\,,\quad
S = \sqrt{1-c^{*}/c}\,,\quad
\theta = 0
\end{equation}
Upon increasing $\alpha_{2}$ and taking $0<\lambda<1$, the system undergoes a transition to a steady state in which $\theta$ and $S$ vary across the system. In addition the flow velocity $v_{x}$ is non-zero and reaches its maximum in the center of the channel. Fig \ref{fig:spontaneous-flow} shows a plot of the hydrodynamic fields as a function of $y/L$. We note that the spatial variations of the order parameter $S$ are not localized, thus this regime is equivalent to the spontaneously flowing state identified in earlier studies of active nematics \cite{Voituriez:2005,Marenduzzo:2007,Giomi:2008,Edwards:2009,Giomi:2010}. It is worth noting that, in this steady state, the nematic order parameter is anti-correlated with the concentration. This feature, which might appear counterintuitive in comparison to the passive case, is a non-equilibrium effect that arises due to the balance between diffusive and active currents (Eq.~\ref{eq:active-current}). Assuming ${\bf n}$ uniform, the total particle current is given by:
\begin{equation}
j_{y}\sim -D\partial_{y}c-\alpha_{1}c^{2}\partial_{y}S
\end{equation}
Since stationary solutions in this geometry require $j_{y}=0$, the active current out of regions with large $S$ is balanced by diffusion from regions with large $c$.

\begin{figure}
\centering
\includegraphics[width=1.\columnwidth]{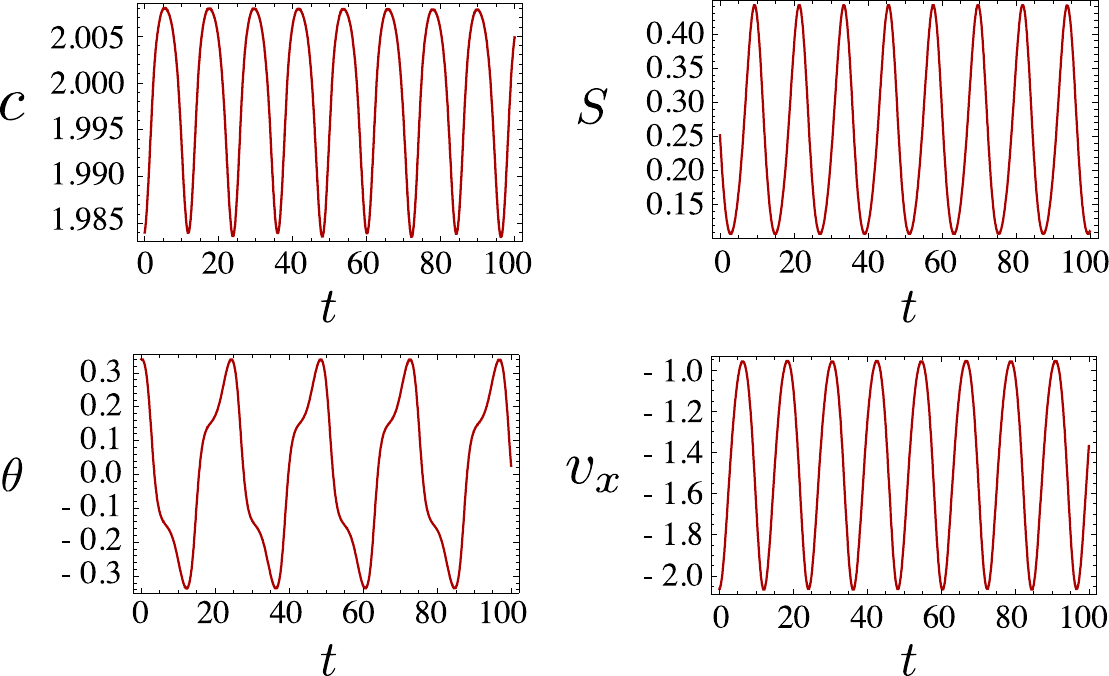}
\caption{\label{fig:oscillatory-flow}The hydrodynamic fields $c$, $S$, $\theta$ and $v_{x}$ in the center of the channel as a function of time in the oscillatory regime obtained by solving Eqs. \eqref{eq:slab-geometry} with boundary conditions \eqref{bc} for $\alpha_{2}=3$ and the other parameters as in Fig. \ref{fig:spontaneous-flow}.}	
\end{figure}

Upon increasing $\alpha_{2}$ the system undergoes a further transition to a regime in which the order parameter $S$, the tilt angle $\theta$, and the velocity oscillate in time, with a frequency that increases with $\alpha_{2}$. In Fig.  \ref{fig:oscillatory-flow} we illustrate the oscillatory behavior by showing a plot of the hydrodynamic fields $c$, $S$, $\theta$ and $v_{x}$ in the center of the channel ($y=L/2$) as a function of time. Both the spontaneously flowing and the oscillatory regime occur in the nematic phase, when the concentration $c>c^{*}$. For $c<c^{*}$, on the other hand, the isotropic homogeneous state with no flow is the only solution.

\subsection{Linear stability analysis}

\begin{figure}
\centering
\includegraphics[width=0.9\columnwidth]{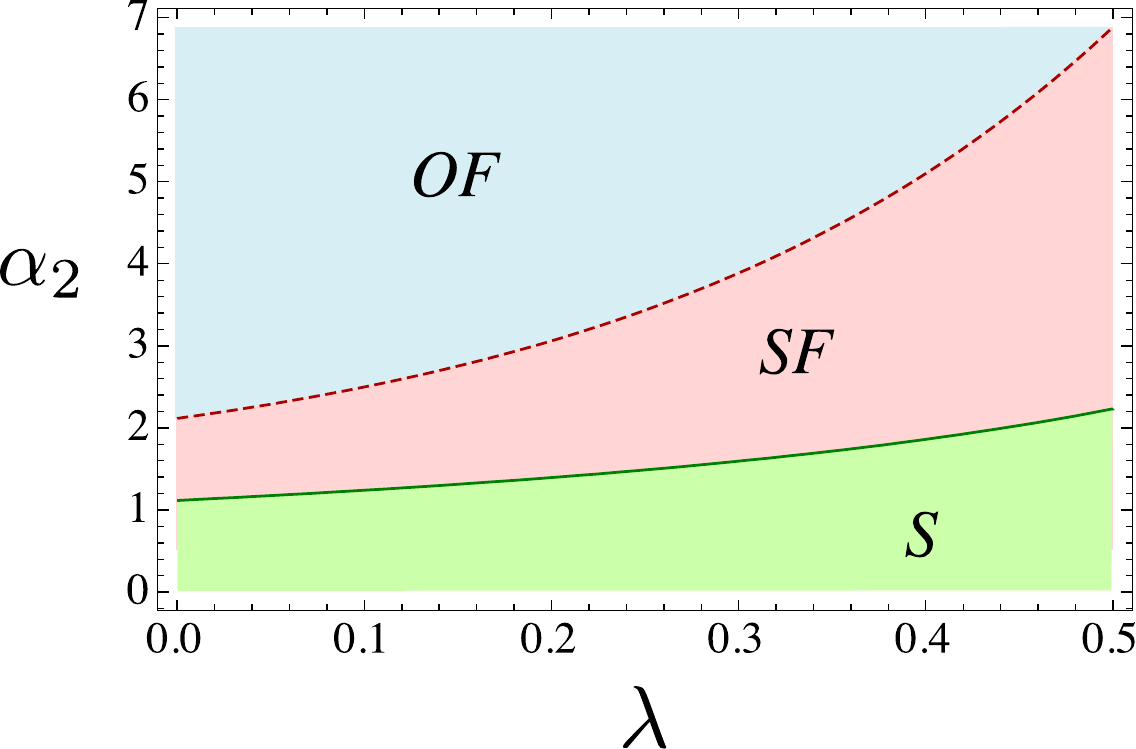}
\caption{\label{fig:phase_diagram}Phase diagram of the flow behavior in the channel geometry, presented in the $(\lambda,\alpha_{2})$ plane (with $\alpha_{1}=\alpha_{2}/10$). The boundary line separating the stationary state ({\em S}) from the steady flow state ({\em SF}) is given by Eq. \eqref{eq:alpha_c1}. The phase boundary of the oscillatory regime ({\em OF}) was obtained numerically. Other parameter values are $\eta=D_{0}=D_{1}=1$, $c_0=2 c^{*}$, and $L=5$.}	
\end{figure}

To understand the result of our numerical simulations and the onset of spontaneous flow we turn to stability analysis of the base state. Letting $\bm{\varphi}=\{c,\,S,\,\theta,\,v_{x}\}$, we consider:
\begin{equation}
\bm{\varphi}(y,t) =  \bm{\varphi}_{0}+\epsilon\bm{\varphi}_{1}(y,t)
\end{equation}
with $\bm{\varphi}_{0}=\{c_{0},\,S_{0},\,0,\,0\}$ the stationary homogeneous solution and $\epsilon\ll 1$. Substituting this ansatz into the hydrodynamic equations \eqref{eq:slab-geometry} yields a linearized system that may be written in block-diagonal form as:
\begin{equation}
\partial_{t}\bm{\varphi}_{1} =
\left(
\begin{array}{cc}
\bm{A} & 0 \\
0 & \bm{B}
\end{array}
\right)	
\bm{\varphi}_{1}
\end{equation}
with:
\begin{equation}
\bm{A} =
\left(
\begin{array}{cc}
 (D_{0}-\tfrac{1}{2}D_{1}S_{0})\,\partial_{y}^{2} &
-\tfrac{1}{2}\alpha_{1}c_{0}^{2}\,\partial_{y}^{2}\\[5pt]
 \tfrac{1}{2}S_{0}(1-S_{0}^{2}) &
-c_{0}S_{0}^{2}+\partial_{y}^{2}
\end{array}
\right)
\end{equation}
and:
\begin{equation}
\bm{B}
=  \left(
\begin{array}{cc}
\partial_{y}^{2} &
-\frac{1}{2}(1-\lambda)\,\partial_{y} \\[5pt]
\alpha_{2} c_{0}^{2}S_{0}\,\partial_{y} &
\eta\,\partial_{y}^{2}	
\end{array}
\right)
\end{equation}
The spontaneous-flow instability is triggered by the coupling between orientation and flow embodied in the $\bm{B}$ operator. To calculate the critical value of $\alpha_{2}$ we must solve the homogeneous system:
\begin{equation}\label{eq:linear-system}
\left\{
\begin{array}{l}
B_{11}\partial_{y}^{2}\theta_{1}+B_{12}\partial_{y}v_{1} = 0\\[5pt]
B_{22}\partial_{y}^{2}v_{1}+B_{21}\partial_{y}\theta_{1} = 0	
\end{array}	
\right.
\end{equation}
with the boundary conditions (\ref{bc}). This implies that the only possible forms for $\theta_{1}$ and $v_{1}$ are:
\begin{subequations}
\begin{gather}
\theta_{1} = C_{1}\sin\left(\frac{2\pi n}{L}\,y\right)\\[5pt]
v_{1} = C_{2}\left[1-\cos\left(\frac{2\pi n}{L}\,y\right)\right]
\end{gather}
\end{subequations}
Substituting these forms into \eqref{eq:linear-system} yields
\begin{equation}
\left\{
\begin{array}{l}
-\frac{(2\pi n)^{2}}{L^{2}}\,C_{1}-\frac{\pi n(1-\lambda)}{L}\,C_{2} = 0\\[5pt]
\frac{2\pi n \alpha_{2} c_{0}^{2} S_{0}}{L}\,C_{1}+\frac{(2\pi n)^{2}\eta}{L^{2}}\,C_{2} = 0	
\end{array}	
\right.
\end{equation}
which together with the requirement for $C_{1}$ and $C_{2}$ to be non-zero yields the following critical value of $\alpha_{2}$:
\begin{equation}
\alpha_{2}^{*} = \frac{8\eta\pi^{2}n^2}{c_{0}^{2}L^{2}S_{0}(1-\lambda)}.
\label{eq:alpha_c1}
\end{equation}
We thus  see the first unstable mode corresponds to $n=1$, which along with the critical value $\alpha_{2}^{*}$ is consistent with that seen in our numerical simulations. The phase-diagram in Fig. \ref{fig:phase_diagram} summarizes the flow behavior for the channel geometry.

\section{\label{sec:plane}Planar geometry}

\subsection{\label{sec:banding}Overview}

We now turn to the case of an active nematic fluid in a two-dimensional square domain with periodic boundary conditions. We numerically integrated the hydrodynamic equations of Sec. \ref{sec:hydrodynamics} using a vorticity/stream-function finite difference scheme on a collocated grid of lattice spacing $\Delta x=\Delta y = 0.078$.  The time integration was performed via a fourth order Runge-Kutta method with time step $\Delta t = 10^{-3}$. As illustrated in Sec. \ref{sec:stability-2d}, the vorticity/stream-function method requires one to solve a Poisson equation at each time step in order to calculate the two components of the flow velocity. This was performed efficiently with a $V$-cycle multigrid algorithm \cite{Multigrid}. As initial configurations we considered a homogeneous system where the director field is aligned along the $x$ axis and subject to a small random perturbation in density and orientation. Thus $c=c_{0}+\epsilon$, $\theta=\epsilon$, $S=\sqrt{1-c^{*}/c}$ and $v_{x}=v_{y}=0$, where $\epsilon$ is a random number of zero mean and variance $\langle \epsilon^{2}\rangle=10^{-2}$. The equations were then integrated from $t=0$ to $t=10^{3}$, corresponding to $10^{6}$ time steps. Except where mentioned otherwise, the numerical calculations described in this section use the parameter values $\alpha_{1}=\alpha_{2}/2$, $\eta=D_{0}=D_{1}=1$, $\lambda=0.1$, $c_0=2c^{*}$ and $L=10$.

\begin{figure}
\centering
\includegraphics[width=0.7\columnwidth]{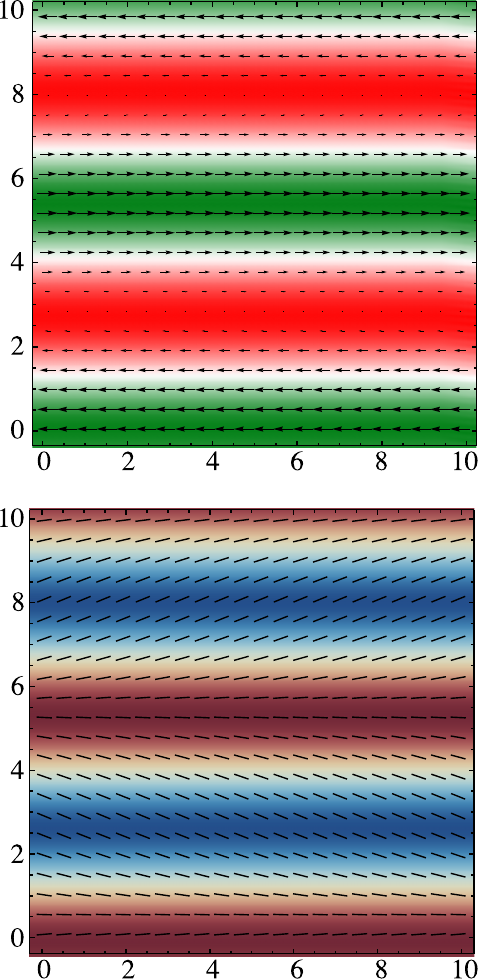}	
\caption{\label{fig:laminar}The velocity field (top) and the director filed (bottom) are superimposed on density plots of the concentration (top) and the nematic order parameter (bottom) for $\alpha_{2}=0.4$ obtained by solving Eqs. \eqref{eq:fullEquations} with periodic boundary conditions. The colors indicate regions of large (green) and small (red) density and large (blue) and small (brown) nematic order parameter. The flow consists of two bands traveling in opposite directions. The director field is nearly uniform inside each band. Parameter values are $c_{0}=2c^{*}$, $\alpha_{1}=\alpha_2/2$, $\lambda=0.1$, $L=10$ and $\eta=D_{0}=D_{1}=1$. For these values, the spontaneous flow transition occurs at $\alpha_{2}^\text{a}=0.37$}
\end{figure}

At low activity, the system relaxes quickly to a stationary homogeneous nematic state with:
\begin{equation}
v_{x}=v_{y}=0\,,\quad
c = c_{0}\,,\quad
S = \sqrt{1-c^{*}/c}\,,\quad
\theta = 0\,.
\end{equation}
Upon raising the activity above a critical value $\alpha_{2}^\text{a}$, with $\alpha_{2}^\text{a}\approx0.37$ for the parameters of our calculation,  this state becomes unstable to a flowing state. The behavior of the spontaneously flowing solution, in this two-dimensional periodic domain, is substantially different than the quasi-one-dimensional system discussed in Sec. \ref{sec:channel}. For values of $\alpha_{2}$ slightly above $\alpha_{2}^\text{a}\approx$ the system divides into two bands flowing in opposite directions. The direction of the streamlines is dictated by the initial conditions which, in this case, favor a flow in the $x$ direction. Moreover the solution is constant along the flow direction (see Fig. \ref{fig:laminar}).

\begin{figure}
\centering
\includegraphics[width=1\columnwidth]{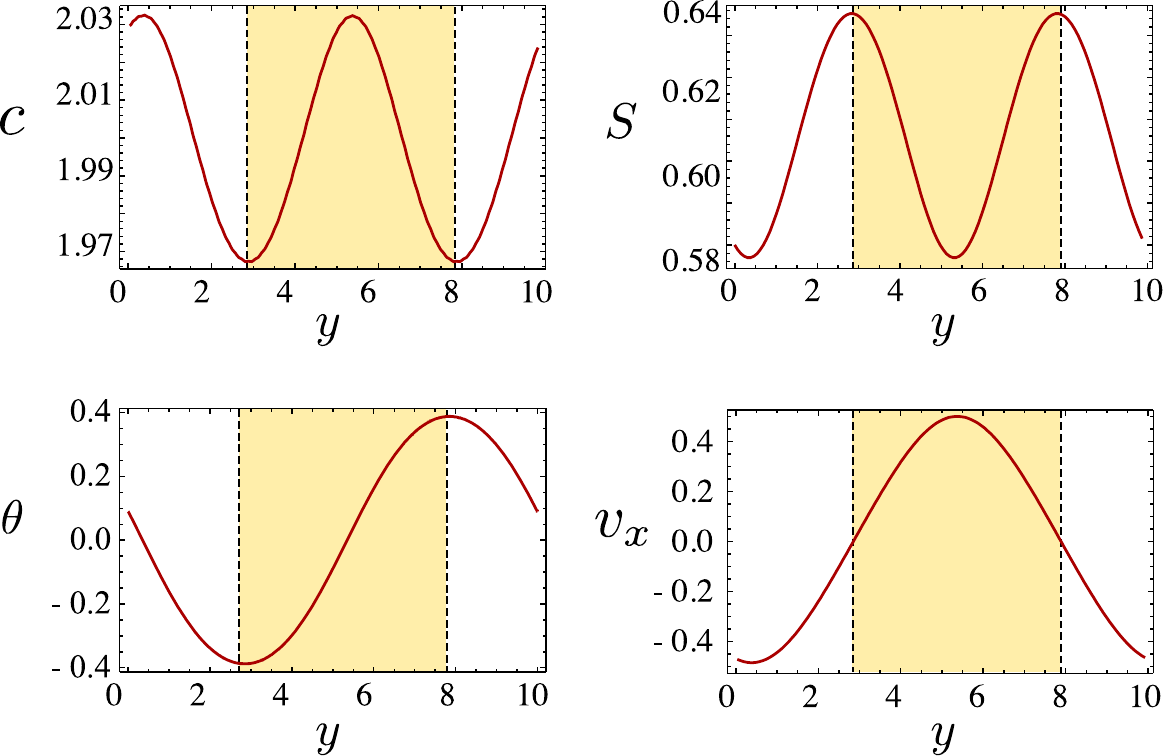}	
\caption{\label{fig:laminar-flow}Hydrodynamic fields $c$, $S$, $\theta$ and $v_{x}$ as a function of $y$ obtained by solving Eqs. \eqref{eq:fullEquations} with periodic boundary conditions. The parameter values are the same as in Fig. \ref{fig:laminar}. The yellow region indicates the extent of a band shown in Fig.~\ref{fig:laminar}.}
\end{figure}

The structure of the bands can be inferred from the plots in Fig. \ref{fig:laminar-flow} showing the various hydrodynamic fields along the $y$ direction. The yellow region indicates the extent of a band. Both the flow velocity and the concentration are maximal at the center of a band. The maximum in the velocity, in particular, is associated with a very sharp variation in the orientation of the director field (see the bottom-left panel of Fig. \ref{fig:laminar-flow}). This rapid variation of the director field generates a large elastic stress, which is balanced by the release of viscous stress through the  increase in the local flow velocity. The nematic order parameter, on the other hand, is minimal in the center of a band due to the balance between diffusive and active currents discussed in Sec. \ref{sec:channel}. As in the case of  spontaneous flow in the channel geometry, here too the variations in concentration and the order parameter are relatively small and not localized.

\subsection{\label{sec:stability-2d}Linear stability analysis}

To understand these behaviors, we analyze the linear stability of the stationary homogenous state in the two-dimensional periodic domain. In order to ensure the incompressibility condition $\nabla\cdot{\bf v}=0$ it is convenient to rewrite the Navier-Stokes equation in terms of vorticity and stream function, by writing:
\begin{equation}
v_{x} = \partial_{y}\psi\qquad
v_{y} =-\partial_{x}\psi
\end{equation}
so that the incompressibility condition is automatically satisfied and the vorticity field is given by:
\begin{equation}\label{eq:vorticity}
\omega=2\omega_{xy}=\partial_{x}v_{y}-\partial_{y}v_{x}
\end{equation}
The two-dimensional Navier-Stokes equation can be expressed in terms of $\omega$ by:
\begin{equation}
\partial_{t}\omega = \eta\Delta\omega+\partial_{x}^{2}\tau_{yx}+\partial_{xy}\tau_{yy}-\partial_{yx}\tau_{xx}-\partial_{y}^{2}\tau_{xy}
\end{equation}
where we defined:
\begin{equation}
\tau_{ij} = -\lambda S H_{ij}+Q_{ik}H_{kj}-H_{ik}Q_{kj}+\alpha_{2}c^{2}Q_{ij}
\end{equation}
From Eq. \eqref{eq:vorticity} we see that the stream-function $\psi$ is related to the vorticity $\omega$ through a Poisson equation of the form: $\Delta\psi = -\omega$. Consistent with the numerical calculations, we consider a nearly uniform suspension of nematogens whose director field is approximatively aligned along the $x$ direction. Thus $c({\bf x},t)=c_{0}+\epsilon\,c_{1}({\bf x},t)$ and ${\bf n({\bf x,t})}={\bf\hat{x}}+\epsilon\,{\bf n}_{1}({\bf x},t)$. Analogously, the nematic tensor can be expressed to first order in $\epsilon$ as:
\begin{equation}
Q_{ij}({\bf x},t) = \frac{S_{0}}{2}(\delta_{ix}\delta_{jx}-\delta_{iy}\delta_{jy})+\epsilon\,Q_{ij}^{(1)}
\end{equation}
with $S_{0}=\sqrt{1-c^{*}/c_{0}}$. As in the quasi-one-dimensional case, we use the compact notation $\bm{\varphi}=\{c,\,Q_{xx},\,Q_{xy},\,\omega\}$ and write the perturbative expression:
\begin{equation}
\bm{\varphi}({\bf x},t) = \bm{\varphi}^{(0)}+\epsilon\,\bm{\varphi}^{(1)}({\bf x},t)
\end{equation}
To enforce periodic boundary conditions on a square domain, we look for solutions of the form:
\begin{equation}
\bm{\varphi}^{(1)}({\bf x},t)
= \sum_{n=-\infty}^{\infty}\sum_{m=-\infty}^{\infty}\bm{\varphi}_{nm}(t)e^{\frac{2\pi i}{L}\,(nx+my)}
\end{equation}
The Fourier components of the stream-function are related to those of the vorticity by:
\begin{equation}
\psi_{nm} = \frac{\omega_{nm}}{\left(\frac{2\pi n}{L}\right)^{2}+\left(\frac{2\pi m}{L}\right)^{2}}
\end{equation}
With this choice the linearized hydrodynamic equations reduce to a set of coupled of linear ordinary differential equations for the Fourier modes $\bm{\varphi}_{nm}$:
\begin{equation}
\partial_{t}\bm{\varphi}_{nm} = \bm{A}_{nm}\bm{\varphi}_{nm}
\end{equation}
with the matrix $\bm{A}_{nm}$ given in Appendix \ref{sec:appendix2}. The first mode to become unstable is the transverse excitation $(n,\,m)=(0,\,1)$ associated with the block-diagonal matrix:
\begin{equation}
\bm{A}_{01} =
\left(
\begin{array}{cc}
\bm{a}_{01} & 0 \\
0 & \bm{d}_{01}
\end{array}
\right)
\end{equation}
with:
\begin{equation}
\bm{a}_{01} =
\left(
\begin{array}{cc}
-\frac{2\pi^{2}}{L^{2}}\,(2D_{0}-D_{1}S_{0}) &
 \frac{4\pi^{2}}{L^{2}}\,\alpha_{1}c_{0}^{2} \\
 \frac{c^{*}}{4c_{0}}\,S_{0} &
-c_{0}S_{0}^{2}-\frac{4\pi^{2}}{L^{2}}
\end{array}
\right)
\end{equation}
and:
\begin{equation}
\bm{d}_{01} =
\left(
\begin{array}{cc}
-\frac{4\pi^{2}}{L^{2}} &
 \frac{1}{2}(1-\lambda)S_{0} \\
 \frac{4\pi^{2}}{L^{2}}\,\alpha_{2}c_{0}^{2}-\frac{16\pi^{4}}{L^{4}}\,S_{0}(1-\lambda) &
-\frac{4\pi^{2}\eta}{L^{2}}
\end{array}
\right)
\end{equation}
The instability first arises from the coupling between local orientations and flow (unless $\alpha_1 \gg \alpha_2$). The critical value of $\alpha_{2}$ is obtained by examining the eigenvalues of the matrix $\bm{d}_{01}$ given by:
\begin{multline}
\Lambda_{\pm}
= -\frac{2\pi^{2}(1+\eta)}{L^{2}}
\pm \frac{\sqrt{2}\,\pi}{L^{2}}\Big[2\pi^{2}(1-\eta)^{2}\\
-4 \pi^2S_{0}^{2}(1-\lambda)^{2}+\alpha_{2}c_{0}^{2}S_{0}L^{2}(1-\lambda)\Big]^{\frac{1}{2}}
\end{multline}
When the real part of the above eigenvalues becomes positive, an instability ensues: this corresponds to  $\alpha_{2}$ larger than the critical value:
\begin{equation}\label{eq:alpha_c2}
\alpha_{2}^\text{a} = \frac{4\pi^{2}[2\eta+S_{0}^{2}(1-\lambda)^{2}]}{c_{0}^{2}L^{2}S_{0}(1-\lambda)}	
\end{equation}
The origin of the instability of the homogeneous stationary state is the same for the 1D channel and the 2D domain and is related to the interplay between the orientation of the director field and the shear flow driven by the internal active stresses. To illustrate this point let us consider a two-dimensional nematic fluid in a stationary state with the director field aligned, say, along the $x$ axis of an arbitrary reference frame. The active stress produced by the action of the motors powers a collective motion of the nematogens. However, the director field rotates in the presence of shear flow for $\lambda\ne 1$, which generates elastic stress. For small activities, the elastic stiffness dominates and suppresses flow, while above the critical value of $\alpha_2^\text{a}$ activity dominates and drives collective motion.  Higher levels of nematic order focus the sources of active stress and thus require lower activity levels to destabilize the homogenous stationary state (lower $\alpha_{2}^\text{a}$).

In a ``dry'' system (i.e. $v_{x}=v_{y}=0$ and $\alpha_{2}=0$) the homogeneous state becomes unstable solely as a consequence of the coupling between density and orientation fluctuations expressed by the matrix $\bm{a}_{nm}$. In this case, the first modes to become unstable are the transverse mode $(1,0)$ and the longitudinal mode $(1,0)$ associated with the matrix (see Appendix A):
\begin{equation}
\bm{a}_{10} =
\left(
\begin{array}{cc}
-\frac{2\pi^{2}}{L^{2}}\,(2D_{0}+D_{1}S_{0}) &
-\frac{4\pi^{2}}{L^{2}}\,\alpha_{1}c_{0}^{2} \\
 \frac{c^{*}}{4c_{0}}\,S_{0} &
-c_{0}S_{0}^{2}-\frac{4\pi^{2}}{L^{2}}
\end{array}
\right)
\end{equation}
Simple algebraic manipulations can be used to show that the real part of the eigenvalues of $\bm{a}_{01}$ and $\bm{a}_{10}$ becomes positive when $\alpha_{1}$ is larger in magnitude than the critical value:
\begin{equation}
\alpha_{1}^{*} = -\frac{2(\pm 2 D_{0}-D_{1}S_{0})(c_{0}L^{2}S_{0}^{2}+4\pi^{2})}{c_{0}c^{*}S_{0}L^{2}}
\end{equation}	
where the plus sign correspond to the $(1,0)$ mode and the minus to the $(0,1)$ mode. This instability, which occurs in absence of hydrodynamics, has been described in various contexts (see for example \cite{Chate:2006,Ginelli:2010} and references therein). We refer the reader to these works for a detailed discussion while in the rest of this article we focus on hydrodynamic phenomena. A thorough discussion on the instability of the homogeneous state in ``dry'' and hydrodynamic systems can be found in \cite{Marchetti:2011}.

\subsection{Relaxation Oscillations}

\begin{figure}
\centering
\includegraphics[width=1.\columnwidth]{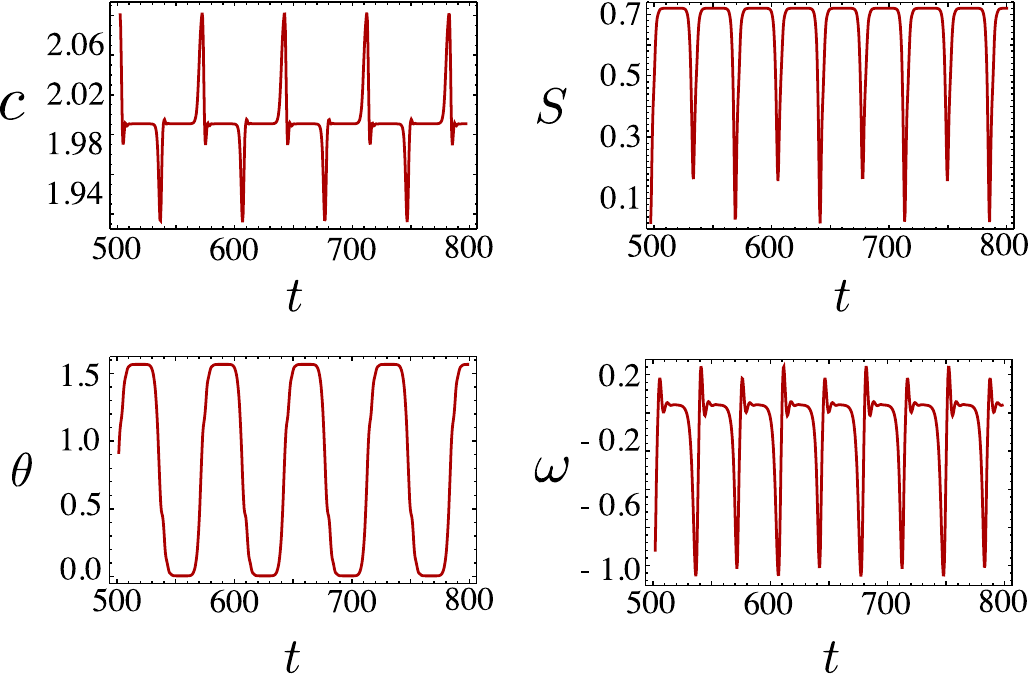}
\caption{\label{fig:flip-flop}Hydrodynamic fields $c$, $S$ $\theta$ and $\omega$ at the center of the box as a function of time obtained by solving Eqs. \eqref{eq:fullEquations} with periodic boundary conditions for $\alpha_{2}=1.5$ and the other parameters as in Fig. \ref{fig:laminar}.}	
\end{figure}

\begin{figure}
\centering
\includegraphics[width=.9\columnwidth]{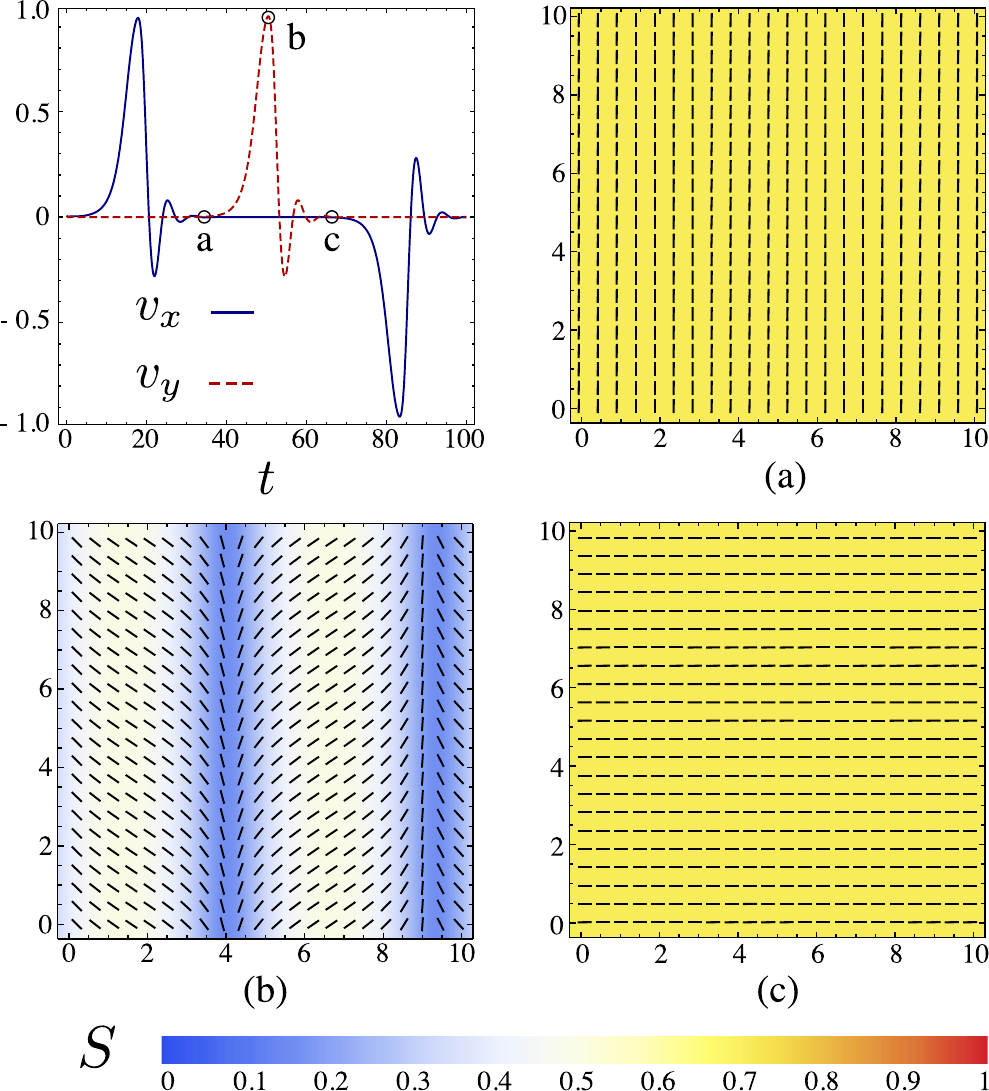}
\caption{\label{fig:burst}Dynamics of an active ``burst'' for the trajectory shown in Fig.~\ref{fig:flip-flop}, with $\alpha_2=1.5$. The flow velocity at the point $x=y=L/3$ is shown as a function of time over the course of a director field rotation (top left) and the director field is shown for the three labeled time points. Between two consecutive bursts the system is homogeneous and uniformly aligned. During a burst, nematic order is drastically reduced in the whole system and the director undergoes a distortion with a consequent formation of two bands flowing in opposite directions. After a burst, a stationary state is restored with the director field rotated of  $90^{\circ}$ with respect to its previous orientation.}		
\end{figure}

Upon increasing the activity parameter $\alpha_{2}$ above a second critical value $\alpha_2^\text{b}$ (with $\alpha_2^\text{b} \approx 0.41$ for our default parameter values), the spontaneously flowing state evolves into a pulsatile spatial relaxation oscillator. Fig. \ref{fig:flip-flop} shows a plot of the various hydrodynamic fields as a function of time for $\alpha_{2}=1.5$. In this regime the dynamics consists of a sequence of almost stationary passive periods separated by active ``bursts'' in which the director switches abruptly between two orthogonal orientations. During passive periods, the particle concentration and the nematic order parameter are nearly uniform across the system, there is virtually no flow, and the director field is either parallel or perpendicular to the $x$ direction. Eventually this configuration breaks down and the director field rotates by 90$^{\circ}$ (see Fig. \ref{fig:burst}). The rotation of the director field is initially localized along lines, generating flowing bands similar to those discussed in Sec. \ref{sec:banding}. The temporary distortion of the director field as well as the formation of the bands is accompanied by the onset of flow along the longitudinal direction of the bands. The flow terminates after the director field rotates and a uniform orientation is restored. The process then repeats.

Remarkably, the rotation of the director fields occurs through a temporary ``melting'' of the nematic phase. As shown in Fig. \ref{fig:flip-flop}, during each passive period the nematic order parameter is equal to its equilibrium value $S_{0}=\sqrt{1-c^{*}/c}$ ($S=1/\sqrt{2}$ because of the choice of $c_{0}=2c^{*}$), but drops to $\sim\frac{2}{5}S_{0}$ during rotation. The reduction of order is system-wide, but, as shown in the bottom-left panel of Fig. \ref{fig:burst}, is most pronounced along the boundaries between bands. Without this transient melting (i.e. if the magnitude of $S$ is not allowed to vary), the distortions of the director field required for a burst are unfavorable for any level of activity.

\begin{figure}
\centering
\includegraphics[width=1\columnwidth]{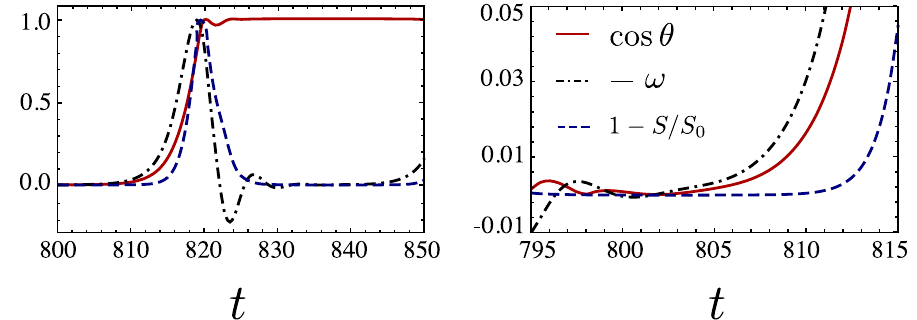}
\caption{\label{fig:timeScales}(Left) The vorticity $\omega$, the orientation of the director $\cos\theta$, and the nematic order parameter $1-S/S_0$ are shown for the point $x=y=L/3$ over the course of a burst for the trajectory shown in Fig.~\ref{fig:flip-flop} with $\alpha_2=1.5$. The data is from the numerical integration and the vorticity is normalized so that its maximum value is one. (Right) A close-up of the same data during the onset of a burst. }	
\end{figure}

A closer look at the dynamics of an individual oscillation elucidates the mechanism of the instability. Fig.~\ref{fig:timeScales} shows the flow (represented by the vorticity), orientation, and the nematic order parameter as a function of time for $\alpha_2=1.5$. Beginning from the homogeneous state, the active forcing generates a gradual increase in flow, and the system evolves in a manner similar to that of the spontaneous flow regime described in section \ref{sec:banding}. As described there, the resulting shear flow causes the nematic director to rotate, generating elastic stress that competes with the active stress. Above the critical value of $\alpha_2$, however, the elastic stress is never sufficient to balance the active stress and the banded flow configuration becomes unstable to melting of the nematic phase. Importantly, the instability occurs only once the flow and director rotation have reached a threshold level; thus, there is a significant delay during which the nematic order parameter is nearly constant. Once melting occurs, the stress is rapidly released during reorientation. The timescale of the oscillation is given by the time required for the flow and director rotation to reach their threshold values, and thus decreases with an increase in  $\alpha_2$ above $\alpha_2^\text{b}$.

\begin{figure}
\centering
\includegraphics[width=1\columnwidth]{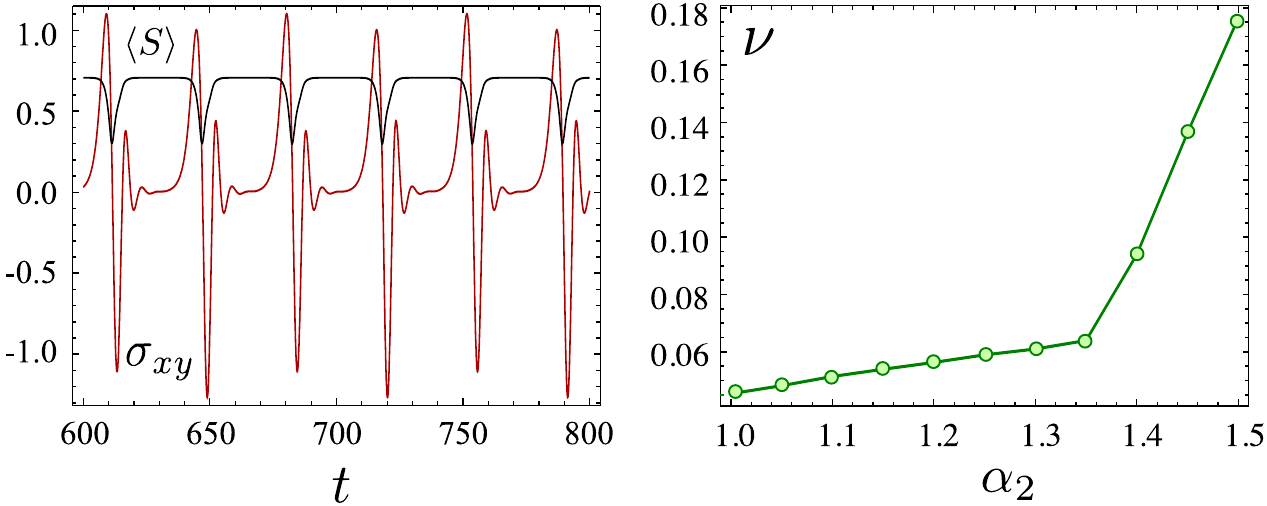}
\caption{\label{fig:stress}(Left) The average nematic order parameter $\langle S\rangle = \int dA/L^{2}\,S({\bf x})$ and the total shear stress $\sigma_{xy}$ are shown over several bursts for for the trajectory shown in Fig.~\ref{fig:flip-flop} with $\alpha_2=1.5$. (Right) The frequency of bursts is shown as a function of $\alpha_{2}$ with other parameters as in Fig.~\ref{fig:flip-flop}.}	
\end{figure}

The physical origin of the oscillatory dynamics in our model of active nematic suspension has to be ascribed to the existence of multiple time scales in a system that is internally driven. As we mentioned in Sec. \ref{sec:hydrodynamics}, one time scale is set by the rate at which the active forcing occurs and is  $\taua=\eta/(\alpha_{2}c_{0}^{2})$. A second time scale is related to the relaxational dynamics of the fluid microstructures (i.e. the solvent flow field, the director field, and the nematic order parameter) and is given by $\taup=\ell^{2}/(\gamma^{-1}K)$ (the time unit in all numerical and analytical calculations). When the two time scales are comparable, the active forcing is accommodated by the microstructures leading to a distortion of the director field and a steady flow. However, when the active forcing occurs at a larger rate the microstructures fail to keep up, revealed above by the instability to melting of the nematic phase. This lag results in oscillatory dynamics and eventually chaos. Similar oscillatory phenomena have been found in models of complex fluids under shear. Cates and coworkers discussed specifically the effect of a slow response of the microstructure to an external shear and showed how such a phenomenon can be naturally described via the FitzHugh-Nagumo equation \cite{Cates:2002,Aradian:2006,Kamil:2010}.

\begin{figure}
\centering
\includegraphics[width=1\columnwidth]{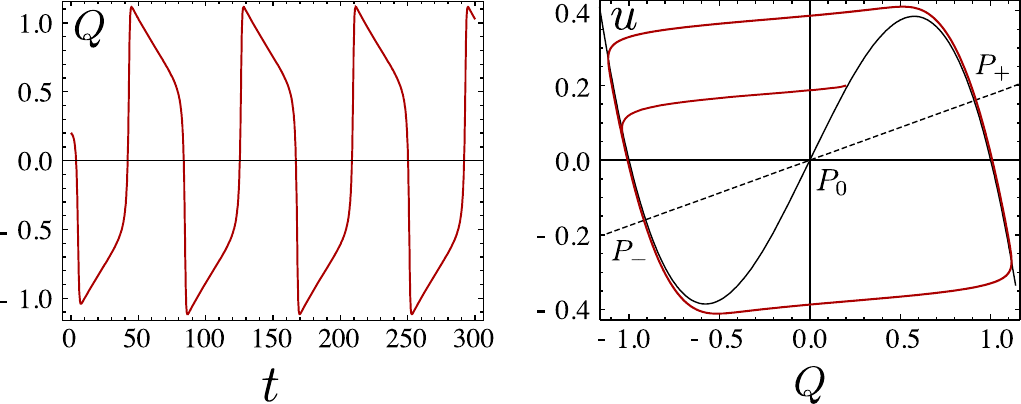}
\caption{\label{fig:nullclines}(Left) A typical trajectory of the variable $Q$ from Eqs. \eqref{eq:fhn2} for $\alpha$ slightly above the critical value $\frac{1}{3}\eta(2a+\eta k^{2})$. (Right) The same limit cycle in the $(Q,u)$-plane. The black dashed line is the $\dot{u}=0$ nullcline and the black solid line is the $\dot{Q}=0$ nullcline.}	
\end{figure}

To illuminate the origins of the relaxation oscillations we construct a simplified version of the hydrodynamic equations that retains the minimal features required to exhibit oscillatory phenomena: the coupling between active forcing and the fluid microstructure and the variable nematic order. The purpose of the following calculation is not to rigorously analyze Eqs.~\ref{eq:fullEquations}, but rather to identify basic physical mechanisms that can drive oscillations and to illustrate the effect of different timescales in the system.

Let us then consider the following simplified version of the hydrodynamic equations for the quantities $Q_{xy}$ and $u_{xy}$ which represent respectively the liquid crystal degrees of freedom and the flow field.
\begin{subequations}\label{eq:fhn1}
\begin{gather}
\dot{Q}_{xy} = u_{xy}+\gamma^{-1}H_{xy}\,, \\[5pt]
\dot{u}_{xy} = \Delta (\eta u_{xy}+\alpha Q_{xy})\,,
\end{gather}
\end{subequations}
obtained by treating $c$ and $Q_{xx}$ as constants and by simplifying the coupling between the nematic tensor and flow, as compared to to the complete phenomenological construction discussed in Sec. \ref{sec:hydrodynamics}. Here, variations in the nematic order parameter are embedded in the Landau-de Gennes free energy within $H_{xy}$. Moving to Fourier space, Eqs. \eqref{eq:fhn1} can be rearranged in the form:
\begin{gather}
\dot{Q}_{xy} = u_{xy}+\gamma^{-1}[(|A|-CQ_{xx}^{2}-k^{2})Q_{xy}-CQ_{xy}^{3}]\,,\notag \\[5pt]
\dot{u}_{xy} =  -k^{2}(\eta u_{xy}+\alpha Q_{xy})\,,
\end{gather}
Finally, by taking $Q=Q_{xy}$, $u=-u_{xy}$, $a=\gamma^{-1}(|A|-CQ_{xx}^{2}-k^{2})$ and $b=\gamma^{-1}C$, one obtains:
\begin{subequations}\label{eq:fhn2}
\begin{gather}
\dot{Q} = aQ-bQ^{3}- u\\[5pt]
\dot{u} =  k^{2}(\alpha Q- \eta u)\,,
\end{gather}
\end{subequations}
equivalent to the spatially homogeneous FitzHugh-Nagumo model or the generalized van der Pol oscillator \cite{Murray,Izhikevich}. In our periodic square domain $k^{2}=(2n\pi/L)^{2}+(2m\pi/L)^{2}$, with $n$ and $m$ integer numbers. The nullclines of the dynamic system \eqref{eq:fhn2} are given by:
\begin{equation}
u=Q(a-bQ^{2}),\qquad u = \alpha Q/\eta
\end{equation}
There are, in general, three fixed points $P=(Q,u)$:
\begin{equation}
P_{0}=(0,0)\,,\quad
P_{\pm}=\left(\pm\sqrt{\frac{a-\alpha/\eta}{b}},\,\pm\frac{\alpha}{\eta}\sqrt{\frac{a-\alpha/\eta}{b}}\right)\,.
\end{equation}
For $\alpha<\eta(2a+\eta k^{2})/3$ the origin $P_{0}$ is a saddle point, while $P_{\pm}$ are stable nodes. A trajectory starting from an arbitrary $(Q,u)$ point will then converge to a stable state characterized by a finite strain-rate that matches the active stress $\alpha Q$: $\eta u=\alpha Q= \alpha\sqrt{(a-\alpha/\eta)/b}$. This fixed point represents the usual spontaneously flowing state (Fig. \ref{fig:laminar}). For $\alpha>\eta(2a+\eta k^{2})$, $P_{\pm}$ become unstable and the system exhibits relaxation oscillations. Fig \ref{fig:nullclines} shows a typical trajectory and a phase-plane plot showing the flow of trajectories in $u$ and $Q$ space. In this regime the dynamics consists of slow relaxations, when the trajectory is close to the cubic nullcline ($\dot{Q}=0$), interspersed with rapid large jumps in $Q$ when the trajectory reaches the unstable portion of the cubic nullcline. By inspection of Eq. (\ref{eq:fhn2}b) the frequency of the oscillations is given by $\nu\sim k^{2}\alpha$. To expand on the assertion that relaxation oscillations arise when the passive timescales exceed that of the active forcing, the critical active rate can be obtained by rewriting  $\alpha_{\rm c}$ in terms of the characteristic timescales defined above as $3 \tau_{\rm a}^{-1} = (2 a \tau_{\rm p}^{-1} + \ell^2 k^2 \tau_{\rm d}^{-1})$, with $\tau_{\rm a}=\eta/\alpha$.

Numerical simulations of the full system (\ref{eq:fullEquations}) exhibits a much richer behavior than that captured by Eqs. \eqref{eq:fhn2}, but the qualitative dependence of the dynamics with respect to $\alpha_2$ persists. The origins of the kink at $\alpha_{2}=1.35$ are unclear at present, but it does not correspond to excitation of a spatial mode of larger wave-number.

\begin{figure}
\centering
\includegraphics[width=0.9\columnwidth]{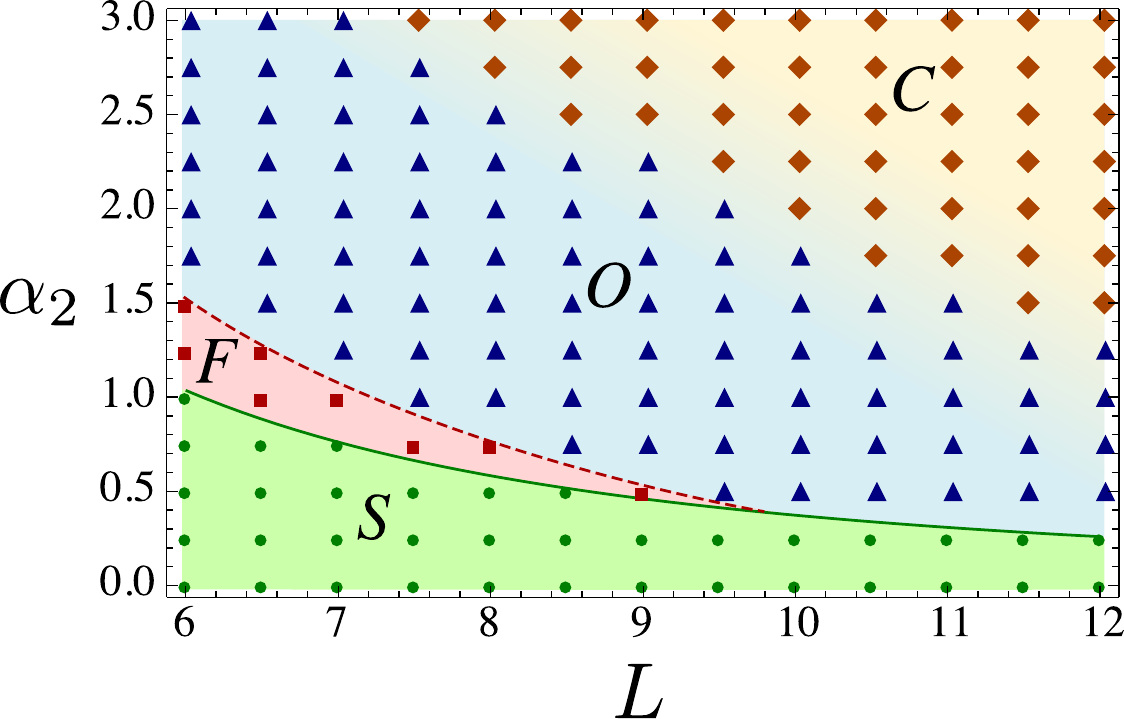}
\caption{\label{fig:phase-diagram2}Phase diagram for the stationary ({\em S}), spontaneous flow ({\em F}), relaxation oscillation ({\em O}) and chaotic ({\em C}) regimes in the plane $(L,\alpha_{2})$ for the full equations Eqs. \eqref{eq:fullEquations} with periodic boundary conditions. The dots are obtained from numerical integration. The green solid line, separating the stationary and flowing state, is given by Eq. \eqref{eq:alpha_c2}. The red dashed line, separating the spontaneously flowing state and the relaxation oscillations regime is interpolated from the numerical data. The color gradient at the intersection between the oscillatory and the chaotic region, indicates a fuzzy boundary between these two regimes. Parameter values are $c_{0}=2c^{*}$, $\alpha_{1}=\alpha_2/2$, $\lambda=0.1$, and $\eta=D_{0}=D_{1}=1$}	
\end{figure}

It is interesting to study how the three regimes described so far change as the size of the system is increased. Fig. \ref{fig:phase-diagram2} shows a phase diagram of the various dynamical regimes for the full equations in the plane $(L,\alpha_{2})$. Upon increasing the size $L$ of the system, the critical value of $\alpha_{2}$ separating the spontaneous flow and the oscillatory regime decreases and merges with the lower phase boundary [whose expression is given in Eq. \eqref{eq:alpha_c2}] for $11<L<12$. Thus we expect that in large samples, the instability of the homogeneous state will lead directly to oscillatory and then chaotic dynamics. The latter is described in the next section.

It is important to emphasize that the excitability described here for active nematic fluids is a purely hydrodynamic phenomenon that arises as a consequence of the existence of multiple time scales in the system, when the dynamics of the flow lags with respect to the rate of the active forcing exerted at the microscopic scale. This phenomenon is thus very different from the large scale fluctuations previously observed in simulations with noise and no hydrodynamics \cite{Chate:2006,Mishra:2006}. Furthermore, the excitability seen here is quite different from that seen in many biological systems where the relaxation oscillations arise from heavily regulated networks of chemical and electrical signals, in contrast with what see in our model where they emerge directly from physical interactions among the constituent components of an active fluid such as the cytoskeleton in a cell.

\subsection{Chaotic regime}

\begin{figure}
\centering
\includegraphics[width=1.\columnwidth]{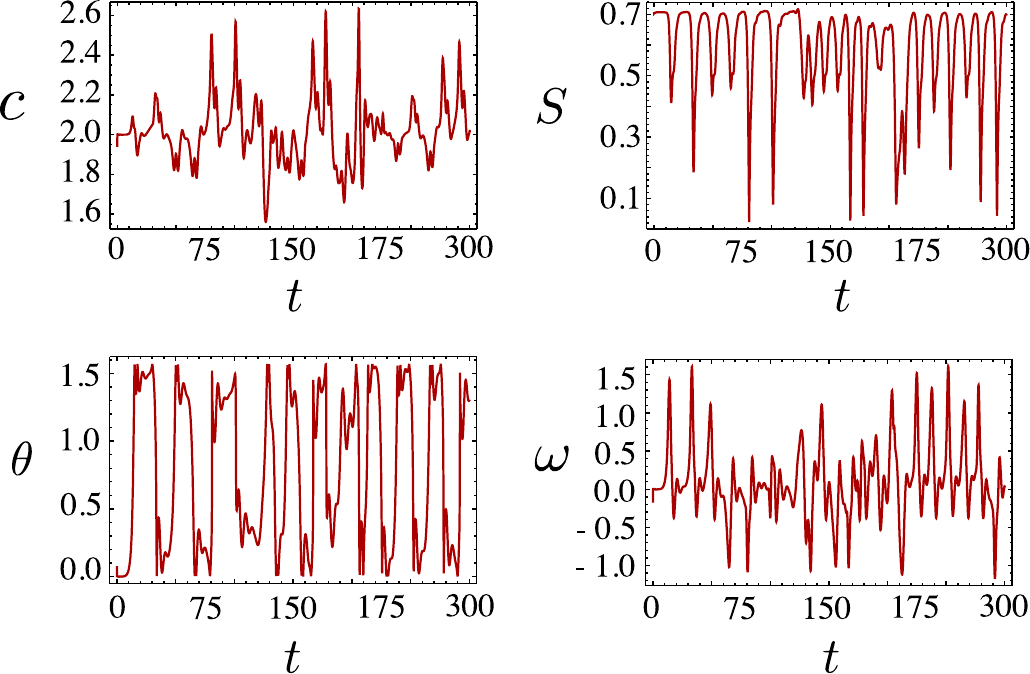}
\caption{\label{fig:chaotic-trajectories1}Hydrodynamic fields $c$, $S$, $\theta$ and $\omega$ at the center of the box as a function of time obtained by solving Eqs. \eqref{eq:fullEquations} with periodic boundary conditions for $\alpha_{2}=2.3$ and the other parameters as in Fig. \ref{fig:laminar}.}	
\end{figure}

When the activity $\alpha_{2}$ is further increased past a third critical value $\alpha_2^\text{c}$, with $\alpha_2^\text{c}\approx2$ for our default parameters (Fig.~\ref{fig:phase-diagram2}), the flow becomes chaotic. The route to chaos takes place through a disordering of the flip-flop dynamics described in the previous section. Initially the dynamics is still characterized by periods of low activity alternating with bursts during which nematic order is temporarily lost and the director field rotates. In Fig. \ref{fig:chaotic-trajectories1} we show the time course of several hydrodynamic fields in a typical trajectory for $\alpha_{2}=2.3$.

\begin{figure}
\centering
\includegraphics[width=0.7\columnwidth]{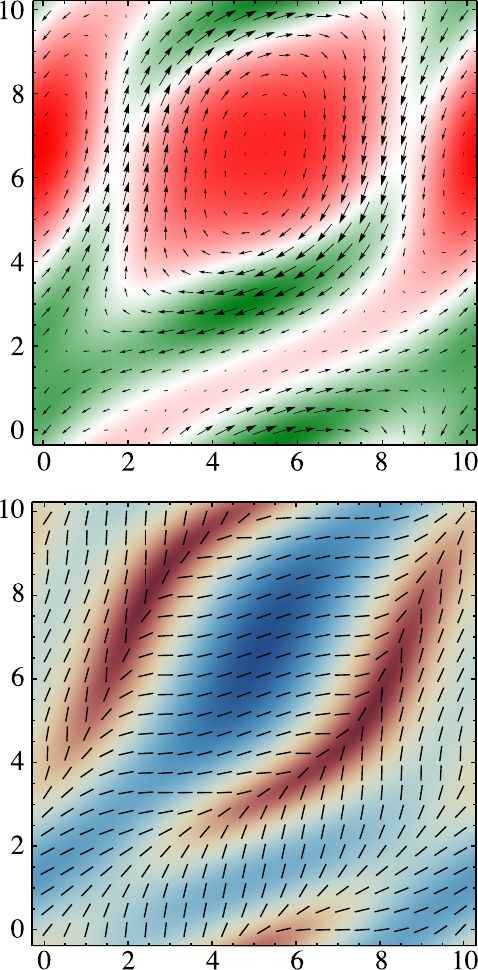}	
\caption{\label{fig:turbulence} (top) The velocity field superimposed on a density plot of the concentration and (bottom) the director field superimposed on a density plot of the nematic order parameter obtained by solving Eqs. \eqref{eq:fullEquations} with periodic boundary conditions for $\alpha_{2}=3$ and other parameters as in Fig.~\ref{fig:laminar}. The colors indicate regions of large (green) and small (red) concentration and large (blue) and small (brown) nematic order parameter.}
\end{figure}

In this chaotic regime, the structure of the flow presents some coherent features  typical of two-dimensional turbulence. For example, in Fig. \ref{fig:turbulence} we show a representative snapshot of the flow velocity superposed on the concentration field, and the director field superposed on the nematic order parameter. We see that the flow is characterized by large vortices that span the system size, with the director field organized into ``grains'' of uniform orientation separated by grain boundaries that span the entire sample. Comparison of the two plots in Fig. \ref{fig:chaotic-trajectories1} reveals that the grain boundaries are the fastest flowing regions in the system. Thus the dynamics in this regime is characterized by grains with approximatively uniform orientation that swirl around each other and continuously merge and reform, giving rise to a flow that appears turbulent. This is similar to other chaotic flows in active fluids that have been reported in models of dilute bacterial suspensions but which do not include liquid crystalline elasticity \cite{Wolgemut:2008,Saintillan:2008b} (also see Ref.~\cite{Marenduzzo:2007} for a related steady state analysis).

\begin{figure}
\centering
\includegraphics[width=0.7\columnwidth]{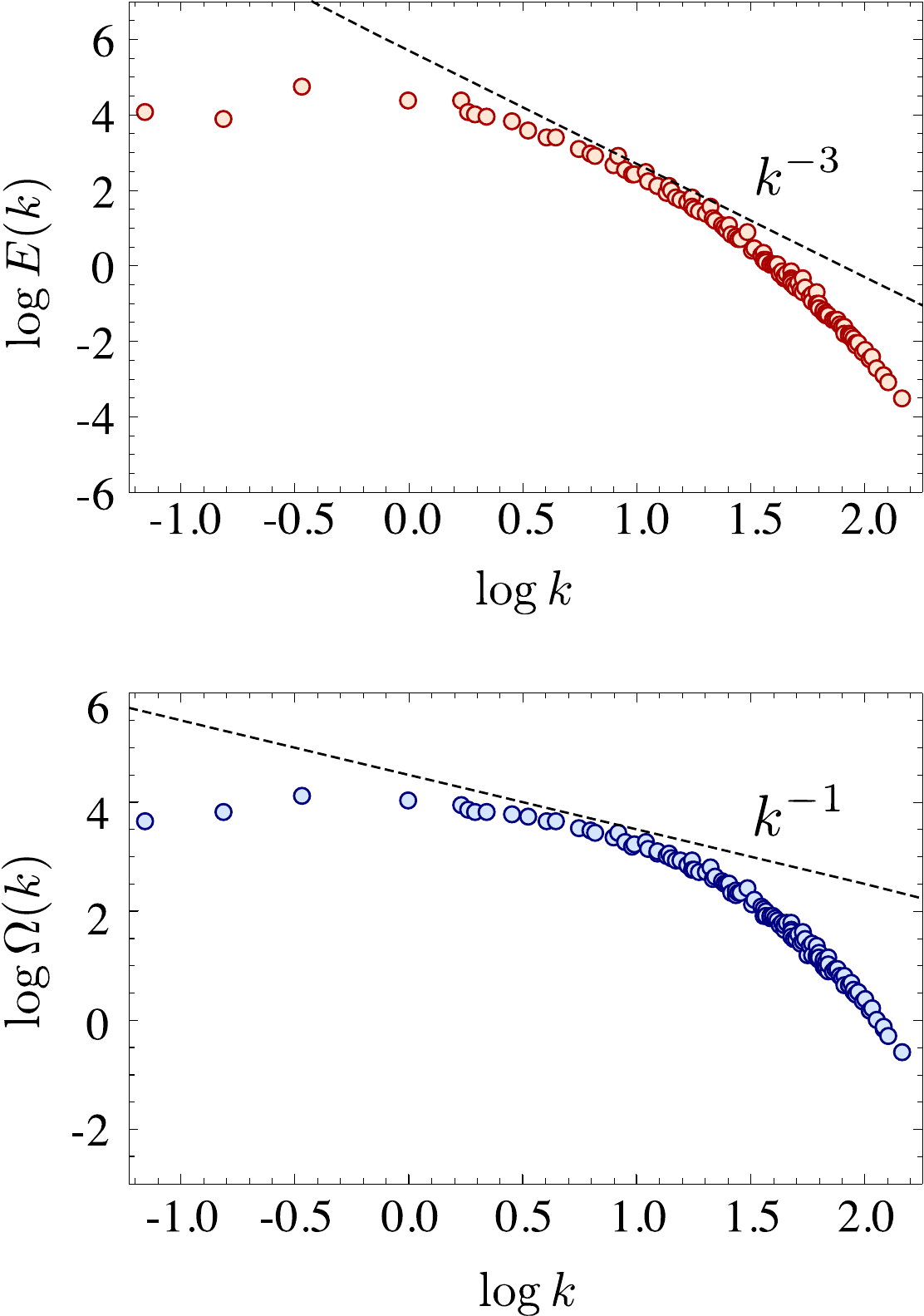}	
\caption{\label{fig:spectra}Energy (top) and enstrophy (bottom) spectra for system of size $L=20$ obtained by solving Eqs. \eqref{eq:fullEquations} for $\alpha_{2}=2$ and other parameters as in Fig. \ref{fig:laminar}. Dashed lines show the graph of the power laws $k^{-3}$ and $k^{-1}$ expected in two-dimensional turbulence.}
\end{figure}

Fig. \ref{fig:spectra} shows the energy and enstrophy power spectra, with the spectral densities $E(k)$ and $\Omega(k)$ defined so that $\frac{1}{2}\langle v^{2} \rangle = \int_{0}^{\infty}dk\,E(k)$ and $\frac{1}{2}\langle \omega^{2} \rangle = \int_{0}^{\infty}dk\,\Omega(k)$ are the mean kinetic energy and enstrophy per unit area.  Although our simple numerical simulations do not span a sufficient range of scales to establish any scaling laws that are expected of two-dimensional turbulence, there are qualitative signatures of such behavior  in our model of active nematic fluids.  We recall that for passive two-dimensional fluids, the classic Kraichnan theory of two-dimensional turbulence in viscous fluids \cite{Kraichnan:1967,Firsch:1996} predicts a double cascade through which energy is transfered from small to large scales while enstrophy flows from large to small scales. At length scales smaller than the injection scale, the enstrophy cascade dominates, giving rise to energy and enstrophy spectra decaying like $k^{-3}$ and $k^{-1}$ respectively (modulo logarithmic corrections). The fundamental difference between simple viscous fluids and the active fluid discussed here is that the forcing acts on a molecular scale here, in contrast with the situation in viscous fluids which is forced at scale of the system. This suggests that a possible mechanism for turbulence in the active fluid described here could involve an \emph{inverse} enstrophy cascade in which vorticity is injected into the system at small scales through the active forcing and then transfered to the scales of order the system size. The dashed lines in Fig. \ref{fig:spectra} show the power laws $E(k)\propto k^{-3}$ and $\Omega(k)\propto k^{-1}$ expected for two-dimensional turbulence in viscous fluids, which while suggestive are not definitive as  our numerical methods are inadequate to stringently test these ideas quantitatively.  However, we hope that our simple discussion might serve as a starting point  for identifying and characterizing {\em active turbulence}.

\section{\label{sec:conclusions}Conclusions and outlook}

In this article we have analyzed in some detail the hydrodynamics of active nematic suspensions in quasi-one and two dimensions. By allowing spatial and temporal fluctuations in the nematic order parameter, we observed a rich interplay between order, activity and flow. Significantly, we find that allowing fluctuations in the magnitude of the order parameter $S$ qualitatively changes the flow behavior as compared to systems in which $S$ is constrained to be uniform.

At a minimal level, the behavior of the system can be qualitatively understood by comparing the timescale of energy input due to activity and the relevant relaxation time scales associated with solvent and liquid crystalline degrees of freedom.  While we have specifically chosen parameter values so that the solvent and liquid crystalline degrees of freedom have the same intrinsic timescales, it would be interesting to continue the analysis to cases with multiple relaxation time scales.

More generally, the richness of behaviors emerging in the present theoretical study of active fluids with liquid crystalline order raises an important question: are these phenomena observed in real active systems ? And if so,  how well can hydrodynamic models capture the complexity of those systems ? In a recent publication Schaller \emph{et al}. \cite{Schaller:2010,Schaller:2010b} reported the observation of many  examples of the collective dynamics in a motility assay consisting of highly concentrated active polar filaments propelled by immobilized molecular motors in a planar geometry. These include the onset of traveling density bands, oscillatory dynamics in which the average orientation of the filaments switches periodically in time, and large scale swirling motions.  Our results suggest that spatially inhomogeneous nematic order is sufficient to drive both an oscillatory dynamics of the director field and a swirling motion even in the absence of polar order.  With this work, we hope to have provided a number of testable predictions that can be used in combination with experiments to shed light on the basic physical mechanisms governing the dynamics of living or otherwise active matter.

\acknowledgments
We gratefully acknowledge support from the Brandeis NSF-MRSEC-0820492(LG, BC, and MFH), the NSF Harvard MRSEC (LG,LM), the Harvard Kavli Institute for Nanobio Science \& Technology (LG, LM), and the MacArthur
Foundation (LM). We thank Cristina Marchetti for useful conversations.

\appendix

\section{\label{sec:appendix1}Nematodynamics via Pauli Matrices}

For some practical application, such as the channel geometry described in Sec. \ref{sec:channel}, it is desirable to have separate hydrodynamic equations for the variables $\theta$ and $S$, rather than having them entangled in the equation for the nematic tensor $Q_{ij}$. In two dimensions, this operation can be performed rather elegantly by using Pauli matrices. To see this let us start from the two-dimensional nematic tensor expressed in matrix form:
\begin{equation}
\bm{Q} = \frac{S}{2}
\left(
\begin{array}{cc}
\cos 2\theta &\phantom{-}\sin 2\theta \\
\sin 2\theta &-\cos 2\theta	
\end{array}
\right)
\end{equation}
In order to decouple $S$ and $\theta$, we can introduce the following matrices:
\begin{subequations}
\begin{gather}
\sigmap = \sin 2\theta\,\bm{\sigma}_{1} + \cos 2\theta\,\bm{\sigma}_{3} \\[5pt]
\bm{\pi} = \cos 2\theta\,\bm{\sigma}_{1}-\sin 2\theta\,\bm{\sigma}_{3}
\end{gather}
\end{subequations}
where $\bm{\sigma}_{1}$ and $\bm{\sigma}_{3}$ are Pauli matrices:
\[
\bm{\sigma}_{1} =
\left(
\begin{array}{cc}
0	&	1\\
1	& 	0	
\end{array}
\right)\,,
\quad
\bm{\sigma}_{2} =
\left(
\begin{array}{cc}
0	&	-i\\
i	& 	0	
\end{array}
\right)
\quad
\bm{\sigma}_{3} =
\left(
\begin{array}{cc}
1	& 0\\
0	&  -1	
\end{array}
\right)
\]
The matrices $\sigmap$ and $\bm{\pi}$ enjoy a number of properties. Namely:
\begin{equation}
\sigmap\,\sigmap = \bm{\pi}\,\bm{\pi} = \bm{\delta}\,,
\qquad\qquad
\sigmap\,\bm{\pi} = i\bm{\sigma}_{2}
\end{equation}
where $\bm{\delta}$ is the $2\times 2$ identity matrix. Since the Pauli matrices are traceless and Hermetian, so are $\sigmap$ and $\bm{\pi}$. An equation for $S$ can be derived straightforwardly by expressing:
\begin{equation}
\bm{Q}= \frac{S}{2}\,\sigmap
\end{equation}
thus:
\begin{equation}
\frac{d\bm{Q}}{dt} = \frac{1}{2}\left(\frac{dS}{dt}\right)\sigmap+S\left(\frac{d\theta}{dt}\right)\bm{\pi}
\end{equation}
Multiplying this expression from the left by $\sigmap$ and taking the trace gives:
\begin{equation*}
\tr\left({\bm \sigmap}\,\frac{d\bm{Q}}{dt}\right) = \frac{1}{2}\left(\frac{dS}{dt}\right)\tr(\bm{\delta}) + iS\left(\frac{d\theta}{dt}\right)\tr(\bm{\sigma}_{2}) = \frac{dS}{dt}
\end{equation*}
Analogously we have that:
\begin{equation}
\tr\left({\bm \pi}\,\frac{d\bm{Q}}{dt}\right) = 2S\left(\frac{d\theta}{dt}\right),
\end{equation}
from which the hydrodynamic equations for $S$ and $\theta$ are found in the form:
\begin{subequations}
\begin{gather}
\frac{dS}{dt}= \tr\left(\sigmap\,\frac{d\bm{Q}}{dt}\right)\\[5pt]
\frac{d\theta}{dt} = \frac{1}{2S}\tr\left(\bm{\pi}\,\frac{d\bm{Q}}{dt}\right)
\end{gather}
\end{subequations}
Thus, the general hydrodynamic equations of Sec. \ref{sec:hydrodynamics} can be finally recast as follows:
\begin{gather}
\partial_{t}{\bf v} = \nabla\cdot\bm{\sigma} \\[5pt]	
[\partial_{t}+{\bf v}\cdot\nabla]S = \left[\lambda_{S}\,{\bm u}+\bm{Q}\bm{\omega}-\bm{\omega}\bm{Q}+\gamma^{-1}{\bm H}\right]_{\sigmap} \notag \\[5pt]	
[\partial_{t}+{\bf v}\cdot\nabla]\theta = \frac{1}{2S}\left[\lambda_{S}\,{\bm u}+\bm{Q}\bm{\omega}-\bm{\omega}\bm{Q}+\gamma^{-1}{\bm H}\right]_{\bm{\pi}} \notag \\[5pt]
[\partial_{t}+{\bf v}\cdot\nabla]c = \nabla\cdot[(D_{0}\bm{\delta}+D_{1}\bm{Q})\nabla c+\alpha_{1}c^{2}\nabla\cdot\bm{Q}] \notag
\end{gather}	
where we used the notation: $[\bm{A}]_{\bm{\alpha}}=\tr[\bm{\alpha}\,\bm{A}]$

\section{\label{sec:appendix2}Linearized System}

In Sec. \ref{sec:stability-2d} we discussed the linear stability of the homogeneous state and we gave an expression for the matrix $\bm{A}_{01}$ of the linearized dynamics associated with the first unstable mode. Here we give an expression for the generic $\bm{A}_{nm}$ matrix. This can be written in the block form:
\begin{equation}
\bm{A}_{nm} =
\left(
\begin{array}{cc}
\bm{a}_{nm} & \bm{b}_{nm} \\
\bm{c}_{nm} & \bm{d}_{nm}
\end{array}
\right)
\end{equation}
with:
\begin{widetext}
\begin{equation}
\bm{a}_{nm} =
\left(
\begin{array}{cc}
-\frac{2\pi^{2}}{L^{2}}[2D_{0}(n^{2}+m^{2})+D_{1}S_{0}(n^{2}-m^{2})] &
\frac{4\pi^{2}}{L^{2}}\alpha_{1}c_{0}^{2}(m^{2}-n^{2}) \\[10pt]
\frac{c^{*}}{4c_{0}}S_{0} &
-\frac{4\pi^{2}}{L^{2}}(n^{2}+m^{2})-c_{0}S_{0}^{2}
\end{array}
\right)
\end{equation}\\[7pt]
\begin{equation}
\bm{b}_{nm} =
\left(
\begin{array}{cc}
-\frac{8\pi^{2}nm}{L^{2}}\,\alpha_{1}c_{0}^{2} &
0 \\[10pt]
0 &
\frac{nm}{n^{2}+m^{2}}\,\lambda S_{0}
\end{array}
\right)
\end{equation}\\[7pt]
\begin{equation}
\bm{c}_{nm} =
\left(
\begin{array}{cc}
0 & 0 \\[10pt]
\frac{2\pi^{2}nmS_{0}}{c_{0}L^{2}}(4\alpha_{2}c_{0}^{2}-\lambda c^{*} S_{0}) &
\frac{8\pi^{2}nm}{L^{2}}[\alpha_{2}c_{0}^{2}-\lambda (c^{*}-c_{0})S_{0}+\frac{4\pi^{2}\lambda S_{0}}{L^{2}}(n^{2}+m^{2})]	
\end{array}
\right)
\end{equation}\\[7pt]
\begin{equation}
\bm{d}_{nm}
= \left(
\begin{array}{cc}
-\frac{4\pi^{2}}{L^{2}}(n^{2}+m^{2}) &
\frac{S_{0}[n^{2}(1+\lambda)+m^{2}(1-\lambda)]}{2(n^{2}+m^{2})} \\[10pt]
\frac{4\pi^{2}}{L^{2}}\alpha_{2}c_{0}^{2}(m^{2}-n^{2})-\frac{16\pi^{4}}{L^{4}}S_{0}(n^{2}+m^{2})[n^{2}(1+\lambda)+m^{2}(1-\lambda)] &
-\frac{4\pi^{2}}{L^{2}}\eta(n^{2}+m^{2})	
\end{array}
\right)
\end{equation}
\end{widetext}

\end{document}